\documentclass[twocolumn,secnumarabic,amssymb,nobibnotes,aps,pra]{revtex4-2}

\usepackage{amsmath}
\usepackage{physics}
\usepackage{graphicx}
\usepackage{xcolor}
\usepackage[justification=RaggedRight]{caption}
\usepackage{subcaption}
\usepackage{hyperref}
\usepackage{cleveref}
\usepackage[inline]{enumitem} 

\begin{document}
\title{
Optical lattice platform for the SYK model
}
\author{Chenan Wei}
\author{Tigran A. Sedrakyan}
\affiliation{Department of Physics, University of Massachusetts, Amherst, Massachusetts 01003, USA}
\date{\today}
\begin{abstract}
The tractability of the Sachdev-Ye-Kitaev (SYK) model at the large $N$ limit makes it ideal to theoretically study its chaotic non-Fermi liquid behavior and holographic duality properties. We show that the complex SYK Hamiltonian emerges from a system of spinless itinerant fermionic atoms in an optical kagome lattice with a strong disorder. We discuss the regimes supporting flat band spectra in a kagome lattice, where the system can be non-dispersive. Random interaction between non-dispersive fermions is induced due to randomly distributed immobile impurities in the optical lattice, that impede the presence of itinerant fermions at their locations. We show that the proposed setup represents a maximally chaotic system and is a reliable experimental platform to realize the SYK model and study its exotic behavior. We show that the velocity distribution of the released fermions is a sensitive probe of the many-body Wigner-Dyson spectral density of states while the averaged many-body Loschmidt echo scheme can measure two-point out-of-time-ordered correlation functions of the SYK system.
\end{abstract}
\maketitle
\section{Introduction}

The Sachdev-Ye-Kitaev (SYK) model, studied by Sachdev and Ye in Ref.~\onlinecite{sachdev1993gapless} and generalized and reexamined recently by Kitaev in Refs.~\onlinecite{Kitaev2015part1,Kitaev2015part2}, has attracted much interest as a strongly interacting system, which exhibits many prominent properties of modern theoretical physics including non-Fermi liquid behavior, AdS/CFT duality, fully chaotic behavior,  and aspects of integrability. The model is an excellent building block for systems where all these properties reveal themselves and lead to fascinating physics that can be studied effectively\cite{davison2017thermoelectric,bagrets2017power,krishnan2017quantum,cotler2017black,gu2017local,klebanov2017uncolored,witten2019syk,garcia2016spectral,bagrets2016sachdev,altland2019syk,fu2018z,SE}. 
The model consists of $N$ fermions with all-to-all random interactions. The Hamiltonian of the model is given by 
\begin{eqnarray}
\mathcal{H}_{\mathrm{SYK}}=-\mu\sum_{i}c_{i}^{\dagger}c_{i}+\sum_{i>j,k>l}J_{ijkl}c_{i}^{\dagger}c_{j}^{\dagger}c_{k}c_{l}, 
\end{eqnarray}
where $i,j,k,l=1\ldots N$, $c_{i}^{\dagger}$($c_{i}$) are fermion creation (annihilation) operators, $\mu$ is the chemical potential, and $J_{ijkl}$ are random couplings with $\left\langle J_{ijkl}\right\rangle=0$ and $\left\langle|J_{ijkl}|^2\right\rangle=\frac{J^2}{2N^3}$. At the large $N$ limit, the SYK model is solvable and its two-point correlation function shows non-Fermi-liquid behavior, and the out-of-time-ordered correlation (OTOC) function displays a maximal Lyapunov exponent\cite{maldacena2016remarks}. At low energy, the SYK model has an emergent conformal symmetry and is dual to an extremal black hole in near AdS$_2$ space\cite{Sarosi2018ads}.

Experimental realization of the SYK model is an important
task, which would allow testing the basic understanding
of the physics behind it. Although the analytical solution shows the duality at the large $N$ and low energy limit (conformal limit), it is still interesting to detect the nearly conformal behavior and its dual black hole in nearly AdS space.
Another motivation for the experimental observation of SYK physics is the following. The SYK model is analytically treatable\cite{Sachdev2015Bekenstein} at $N\gg1$, and large time scales $J\tau\gg1$, where the system is ergodic, conformally invariant, and supports the universal many-body Wigner-Dyson statistics. At small time scales, the system is non-ergodic. The time scale at which the ergodicity sets in, called Thouless time, is outside of the solvable limit. The Thouless time (or the corresponding Thouless energy) and its scaling with $ N $ are so far unknown. It would be fascinating to experimentally observe the system behavior near and beyond the Thouless energy. Thus a cold atom experiment, if realized, would undoubtedly be invaluable and would shed light on these issues.

There is already activity in this direction. Several possible realizations of SYK model in experiments were proposed, such as the Majorana SYK model at the interface of a topological insulator and superconductor\cite{pikulin2017black}, the Majorana SYK model with a quantum dot coupled to an array of topological superconducting wires\cite{chew2017approximating}, the real SYK model in an optical lattice loaded with atoms and molecules\cite{danshita2017creating}, and the complex SYK model in a graphene flake in a magnetic field\cite{chen2018quantum}.  Another avenue for studies of the model is the digital quantum simulation proposed in Ref.~\onlinecite{Garcia-Alvarez2017Digital}. Following this idea, a generalized SYK-like model is studied in Ref.~\onlinecite{luo2019quantum}.

A vital ingredient of the SYK model is that it is zero space dimensional. Therefore, in order to fabricate the SYK model experimentally in real $d$-dimensional space, it is necessary to eliminate the momentum dependence of the spectrum and make a flat band. In Ref.~\onlinecite{chen2018quantum}, this problem is solved
by introducing a strong magnetic field, which forces energy levels of the system to become flat Landau levels. Besides the flatness of the band, one also needs to generate random couplings for interactions of fermions. In Ref.~\onlinecite{chen2018quantum}, this problem is solved by considering an ensemble of samples with random boundary conditions.

In this paper, we present another scheme of flat band formation and randomization of the interaction coupling. We propose a concrete realistic realization of the SYK model with cold atoms in an optical lattice of kagome type, which gives straightforward ways of detecting its nontrivial properties experimentally.  Among those is the measurement of the distribution of the particle velocities after deconfinement of the optical lattice. It will manifest the many-body Wigner-Dyson spectral density of states. The second possibility is the measurement of the averaged two-point OTOC function. The kagome optical lattice realized in Ref.~\onlinecite{jo2012ultracold,leung2020interaction} shows a flat band\cite{leykam2018artificial}, which is an ideal playground for studying enhanced interaction effects of particles. In this paper, we show that the low energy effective theory of spinless fermions in an optical kagome lattice with strong disorder realizes the complex SYK model. Unlike previous proposals, this method does not need superconductors or strong magnetic fields, while the disorder can be tuned. 

The remainder of the paper is organized as follows. In \cref{model}, we discuss the effective theory and the proposed experimental setup. In \cref{SYK}, we show that the low energy physics of the proposed setup is dominated by the SYK model. Finally, \Cref{experiment} presents estimates for the experimental realization of the proposed setup.

\section{The model and the proposed setup}\label{model}

\begin{figure}
\begin{subfigure}{0.5\textwidth}
\includegraphics[width=0.49\textwidth]{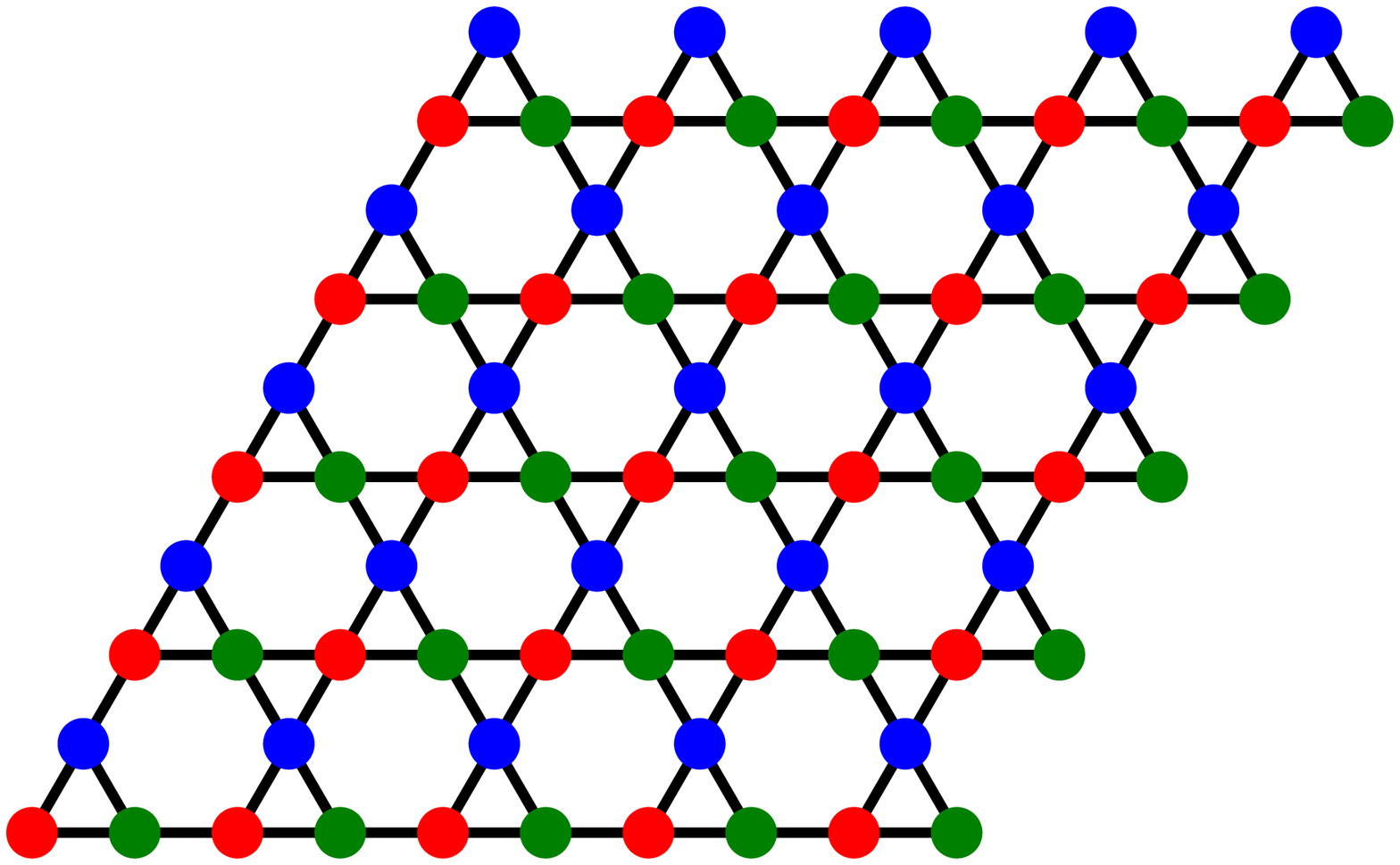}
\includegraphics[width=0.49\textwidth]{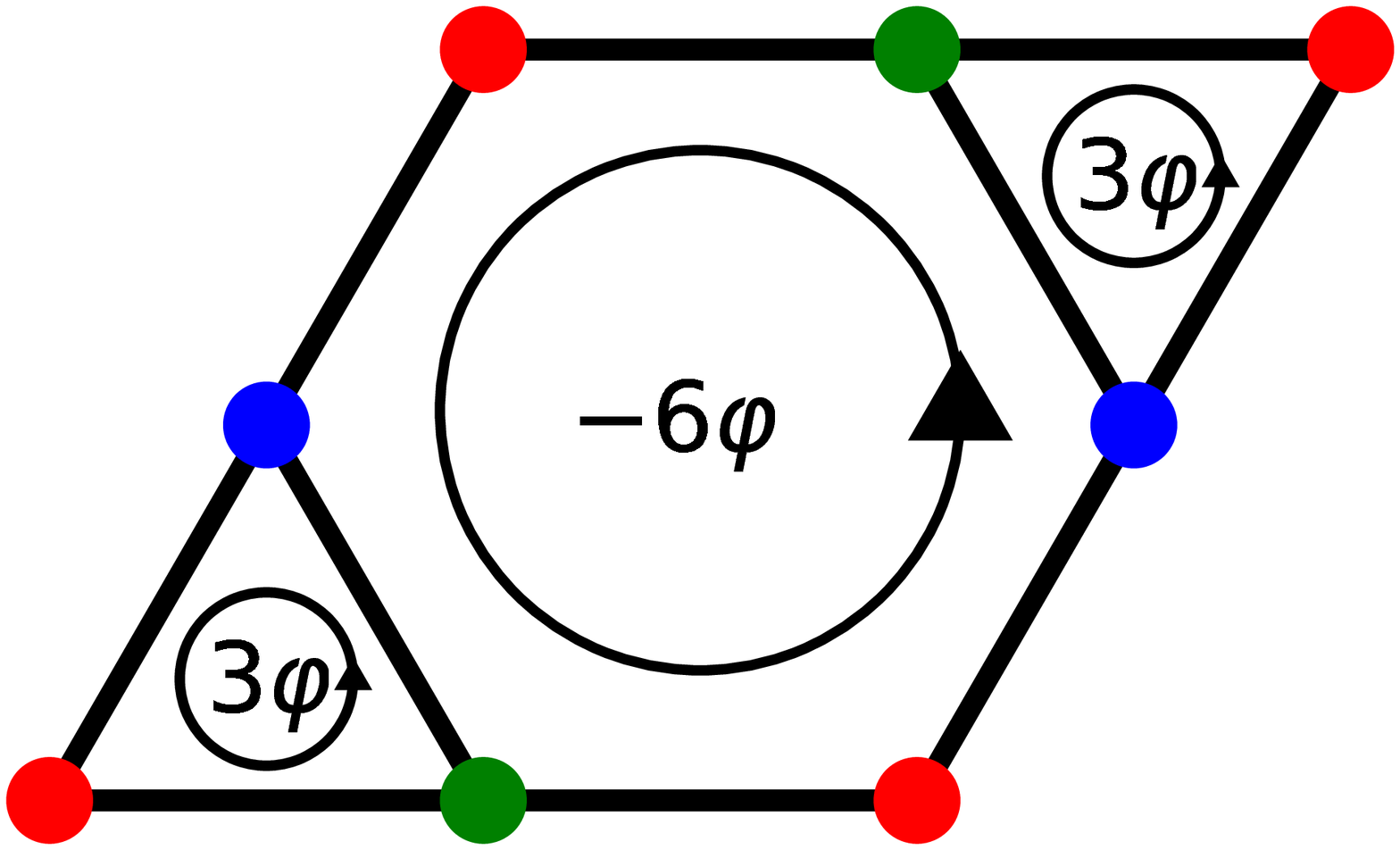}
\begin{picture}(0,0)
\put(-130,100){$(a)$}
\put(-39,96){\line(1,0){19.5}}
\put(-48,78){\line(1,0){19.5}}
\put(-48,78){\line(1,2){9}}
\put(-28.5,78){\line(1,2){9}}
\put(-23,86){\vector(1,0){65}}
\end{picture}
\phantomsubcaption{\label{lattice}}
\end{subfigure}
\begin{subfigure}{0.15\textwidth}
\includegraphics[trim={5 10 5 5},clip,width=\textwidth]{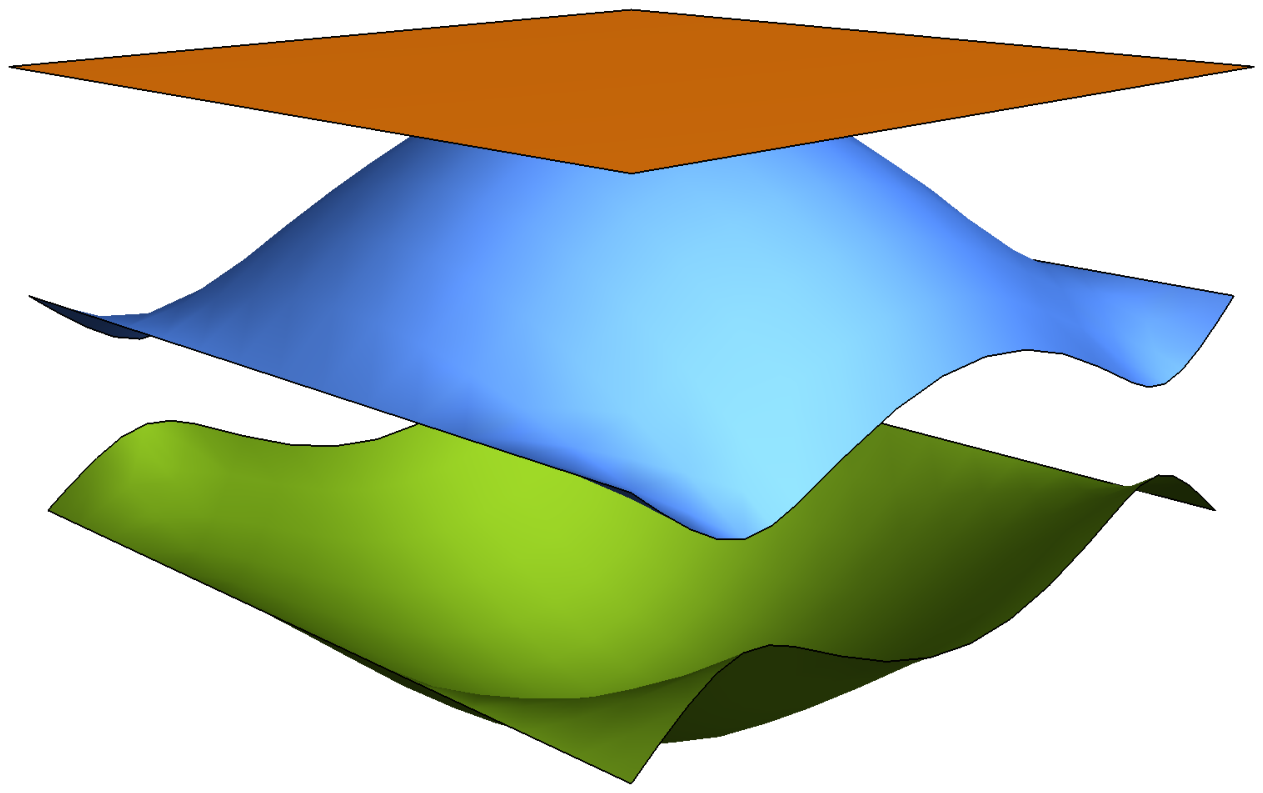}
\begin{picture}(0,0)
\put(-45,80){$(b)$}
\end{picture}
\phantomsubcaption{\label{Pi0}}
\end{subfigure}
\begin{subfigure}{0.15\textwidth}
\includegraphics[trim={5 5 5 5},clip,width=\textwidth]{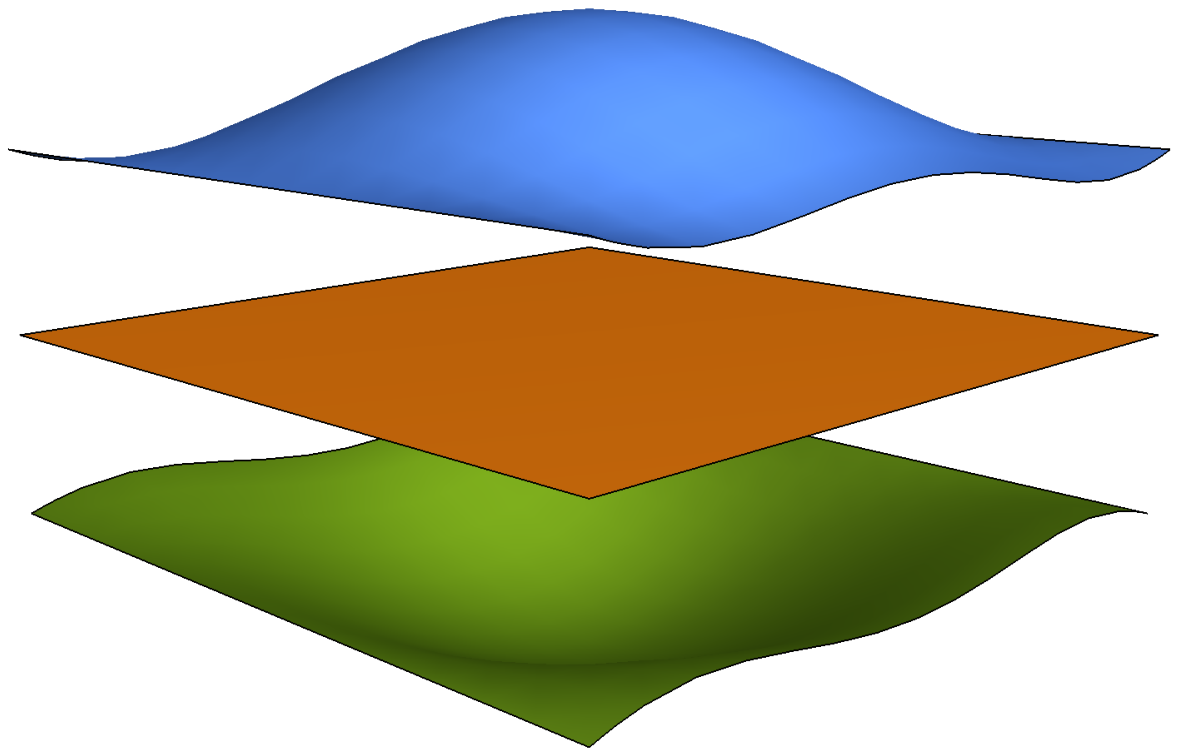}
\begin{picture}(0,0)
\put(-45,80){$(c)$}
\end{picture}
\phantomsubcaption{\label{Pi6}}
\end{subfigure}
\begin{subfigure}{0.15\textwidth}
\includegraphics[trim={5 5 5 10},clip,width=\textwidth]{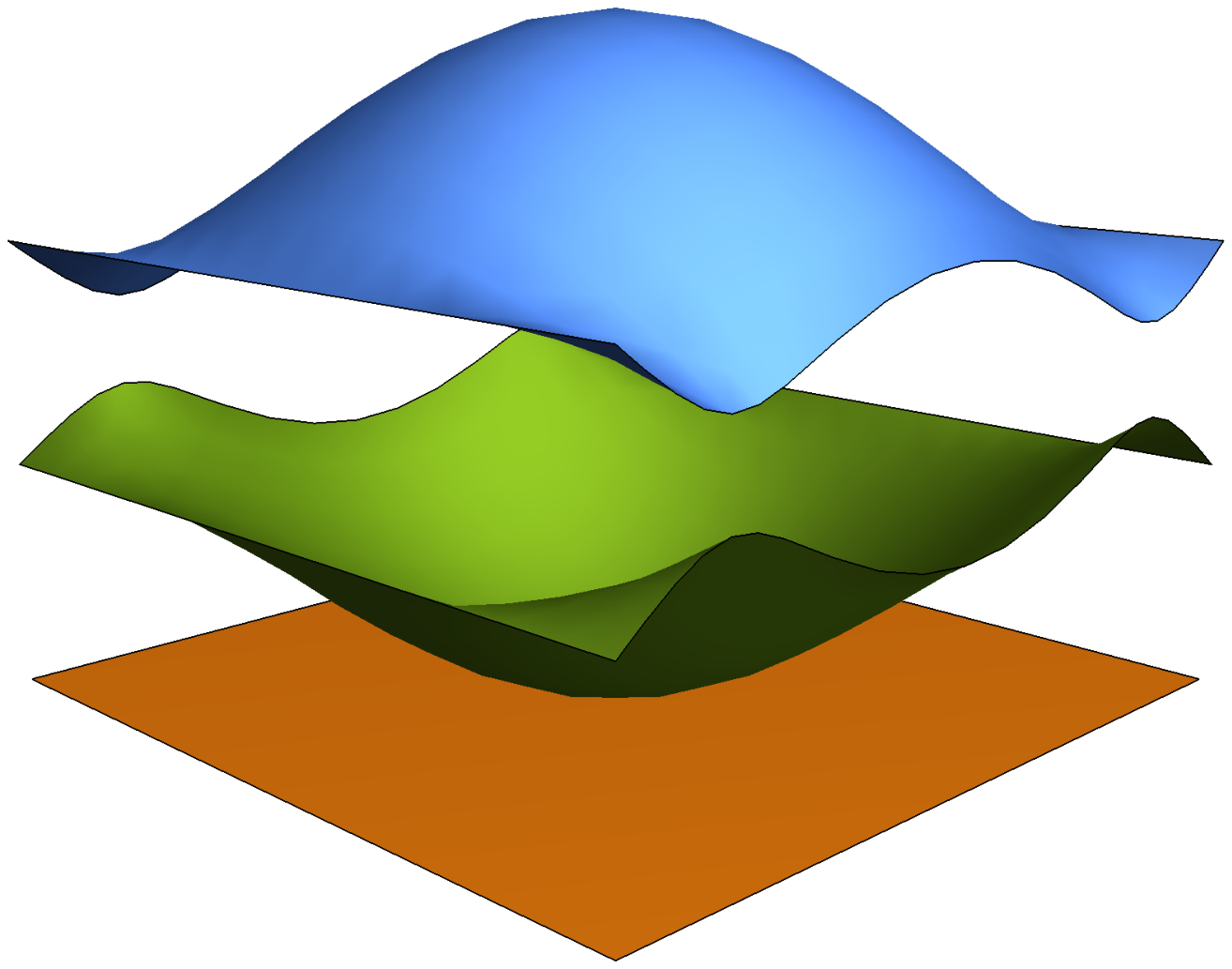}
\begin{picture}(0,0)
\put(-45,80){$(d)$}
\end{picture}
\phantomsubcaption{\label{Pi1}}
\end{subfigure}
\vspace{-5mm}
\caption{(Color online). (a) A fragment of the kagome optical lattice with nearest-neighbor couplings. A unit cell of the lattice is shown with the corresponding flux configuration that supports a flat band. The bare band spectrum with a flat band is shown for $\varphi=0$ (b), $\varphi=\pi/6$ (c), and $\varphi=\pi$ (d). \label{kagome}}
\end{figure}

\begin{figure}
\begin{subfigure}{0.23\textwidth}
\includegraphics[trim={5 10 5 10},clip,width=\textwidth]{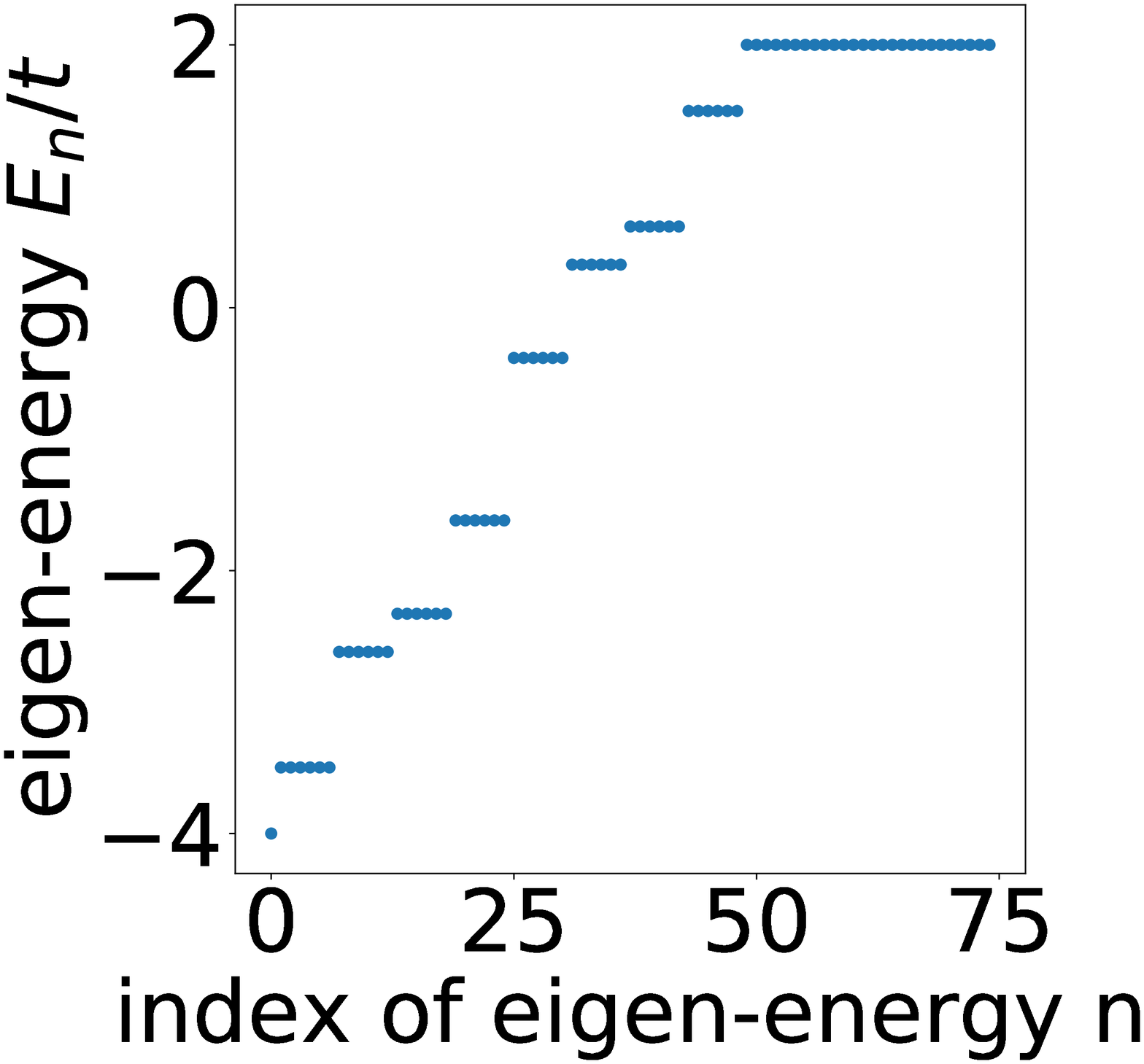}
\begin{picture}(0,0)
\put(-22,108){$(a)$}
\end{picture}
\phantomsubcaption{}
\vspace{-3mm}
\end{subfigure}
\begin{subfigure}{0.23\textwidth}
\includegraphics[trim={5 10 5 10},clip,width=\textwidth]{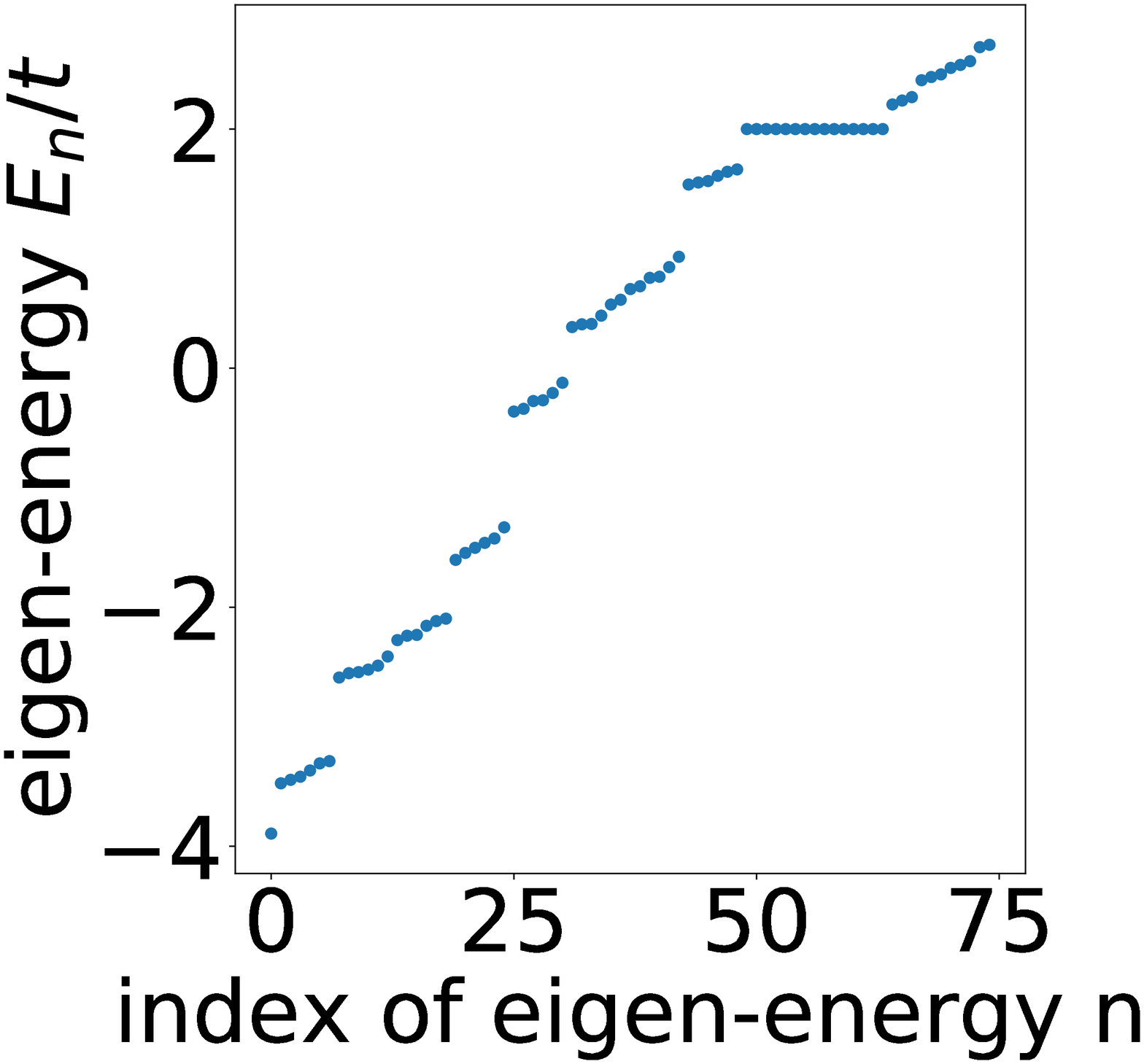}
\begin{picture}(0,0)
\put(-22,108){$(b)$}
\end{picture}
\phantomsubcaption{}
\vspace{-3mm}
\end{subfigure}
\begin{subfigure}{0.23\textwidth}
\includegraphics[trim={5 10 5 10},clip,width=\textwidth]{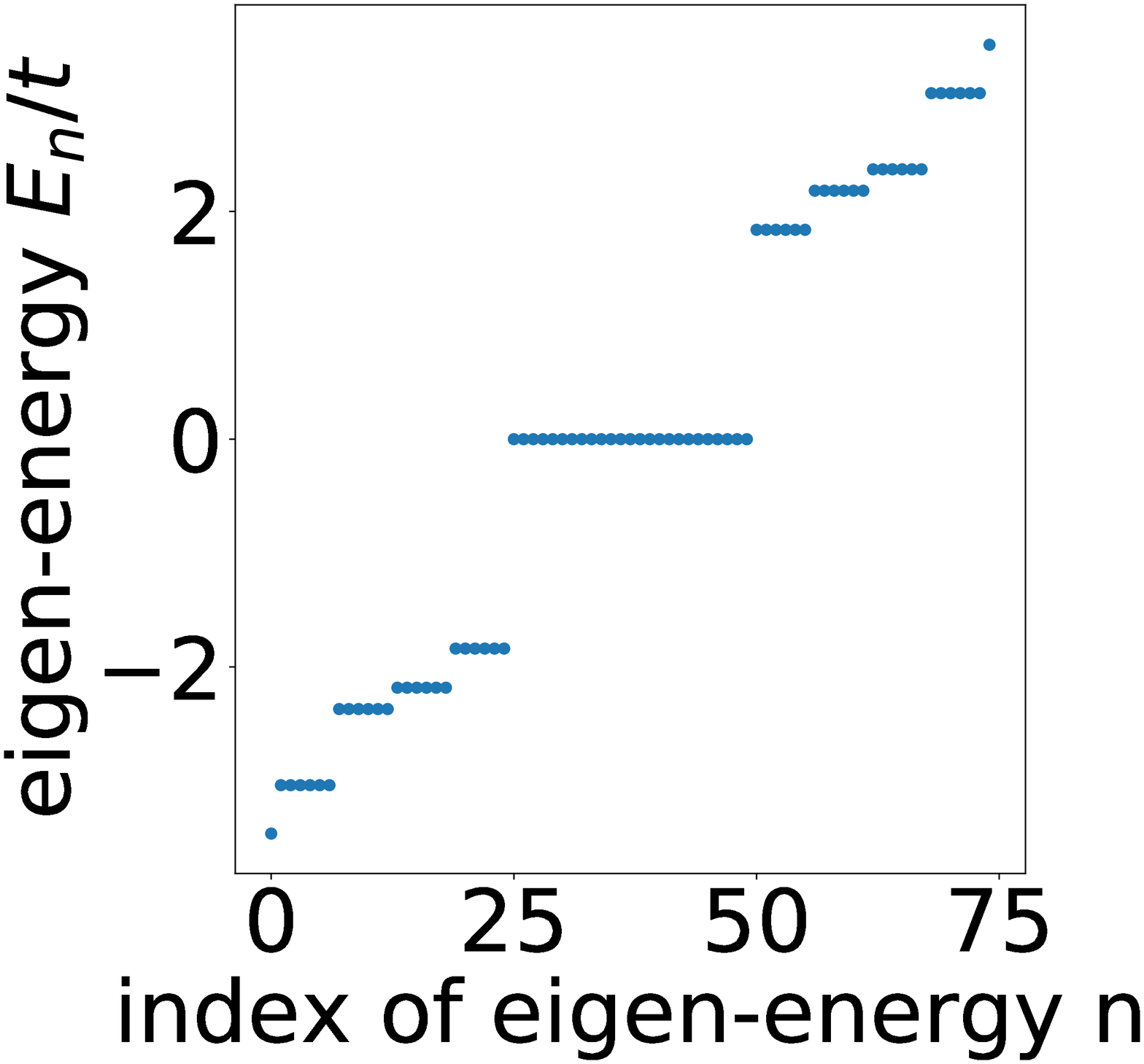}
\begin{picture}(0,0)
\put(-22,108){$(c)$}
\end{picture}
\phantomsubcaption{}
\vspace{-3mm}
\end{subfigure}
\begin{subfigure}{0.23\textwidth}
\includegraphics[trim={5 10 5 10},clip,width=\textwidth]{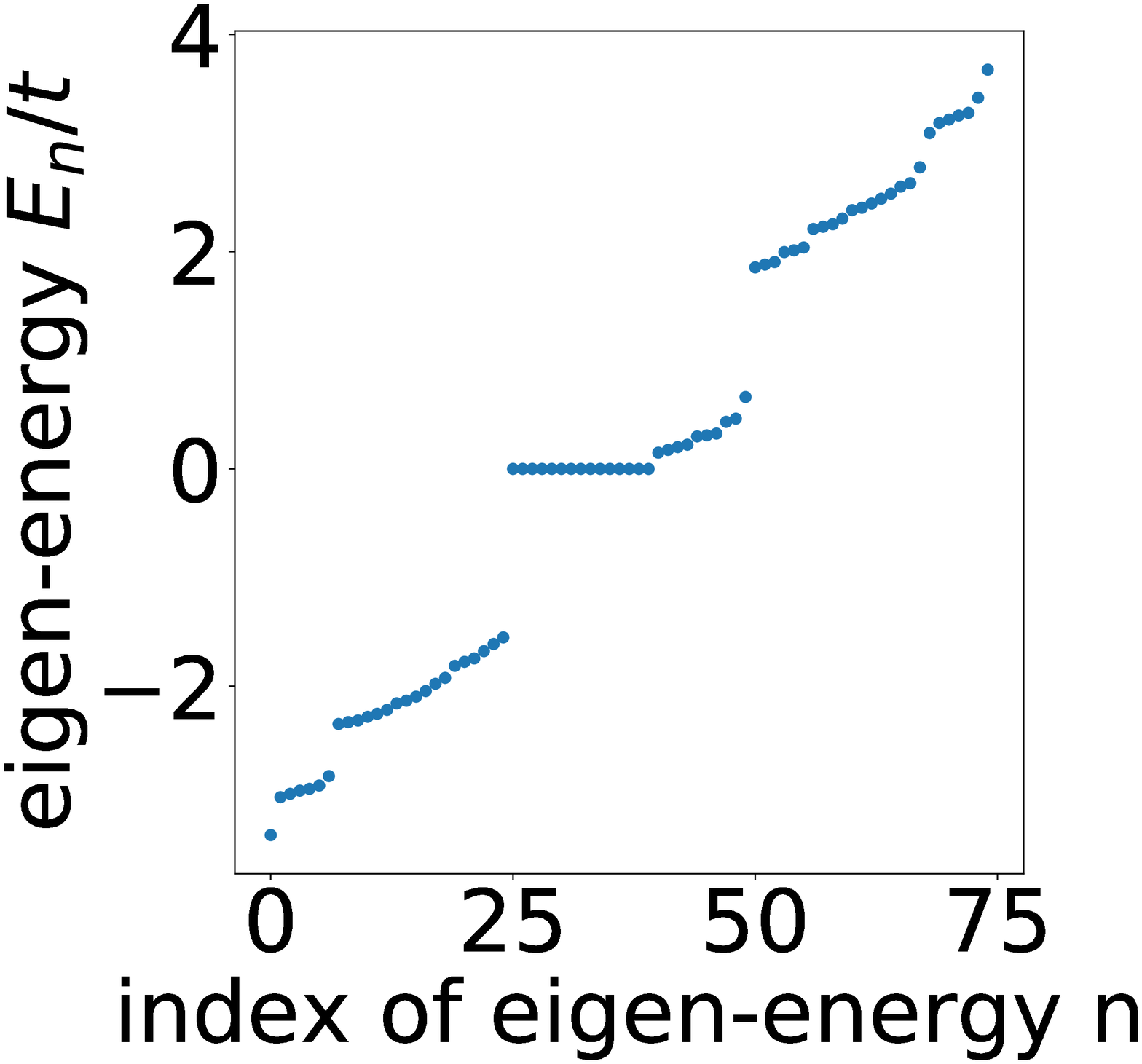}
\begin{picture}(0,0)
\put(-22,108){$(d)$}
\end{picture}
\phantomsubcaption{}
\vspace{-3mm}
\end{subfigure}
\begin{subfigure}{0.23\textwidth}
\includegraphics[trim={5 10 5 10},clip,width=\textwidth]{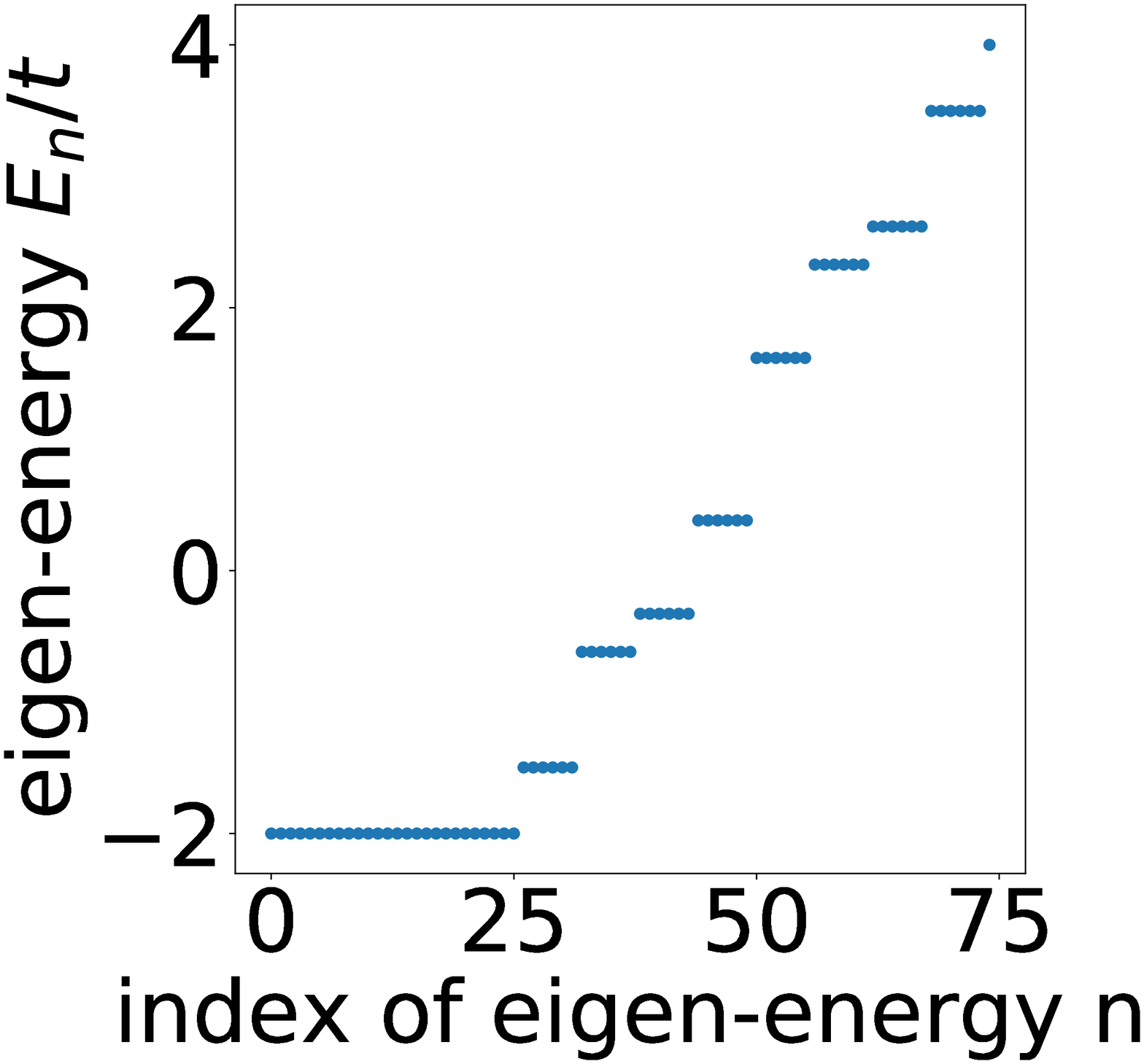}
\begin{picture}(0,0)
\put(-22,108){$(e)$}
\end{picture}
\phantomsubcaption{}
\end{subfigure}
\begin{subfigure}{0.23\textwidth}
\includegraphics[trim={5 10 5 10},clip,width=\textwidth]{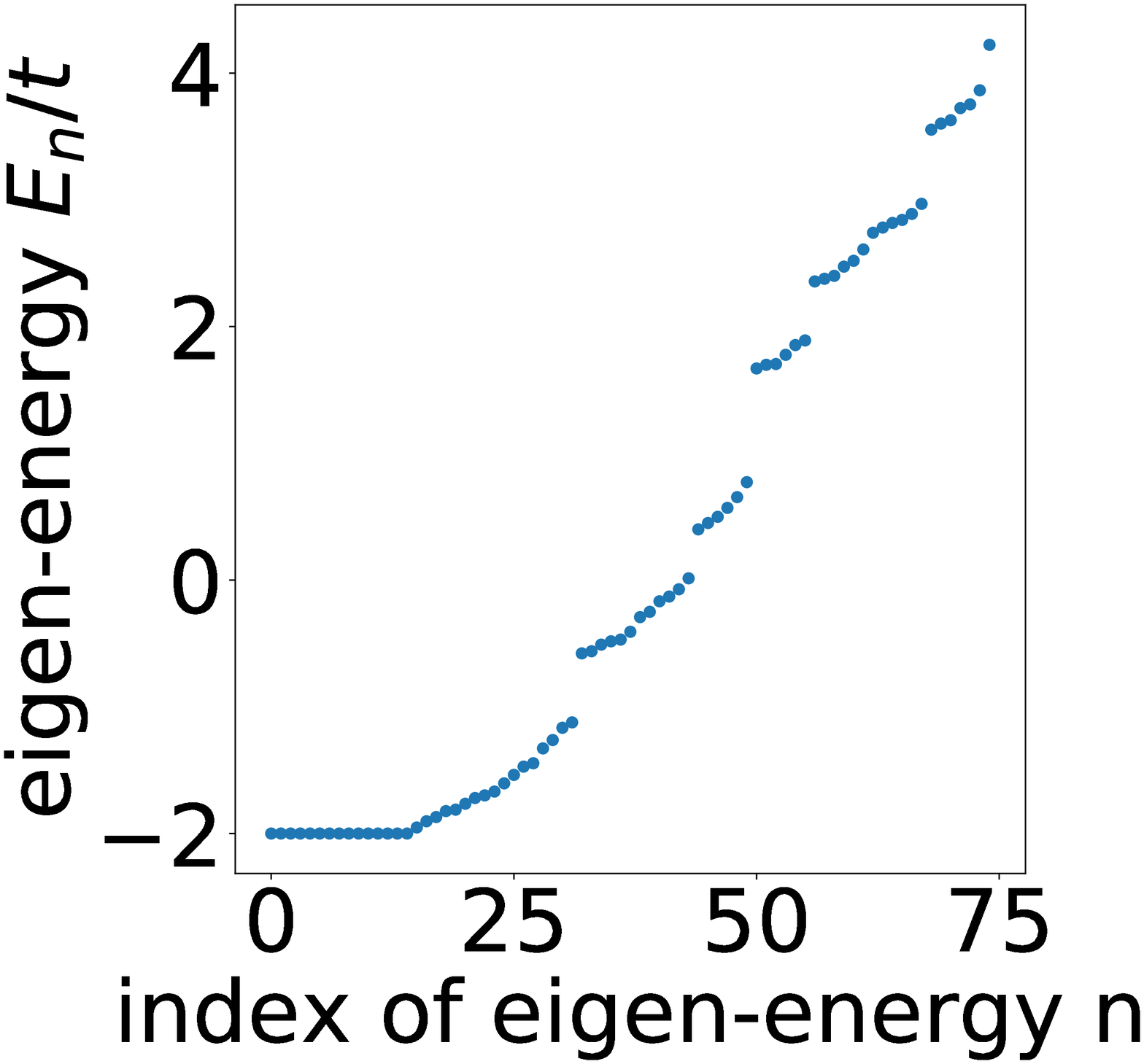}
\begin{picture}(0,0)
\put(-22,108){$(f)$}
\end{picture}
\phantomsubcaption{}
\end{subfigure}
\vspace{-2mm}
\caption{(Color online). The exact spectrum of the non-interacting system plotted from the Hamiltonian $\mathcal{H}_{0}$ (left column) and $\mathcal{H}_{0}+\mathcal{H}_{imp}$ (right column) with 25 unit cells. The spectrum is shown for $\varphi=0$ in the absence of impurities (panel (a)) and in the presence of impurities (panel (b)); for $\varphi=\frac{\pi}{6}$ in the absence of impurities impurities (panel (c)) and in the presence of impurities (panel (d)); for $\varphi=\pi$ in the absence of impurities (panel (e)) and in the presence of impurities (panel (f)).\label{spectrum}}
\end{figure}

We now concentrate on the details of the Hamiltonian describing the proposed scheme. It consists of three terms: $\mathcal{H}=\mathcal{H}_{0}+\mathcal{H}_{imp}+\mathcal{H}_{int}$, with the tight-binding Hamiltonian $\mathcal{H}_{0}$ of itinerant ultracold fermions supporting a flat band,  the impurity Hamiltonian $\mathcal{H}_{imp}$ describing a set of randomly distributed on-site $\delta$-function potentials, and a short-range interaction Hamiltonian $\mathcal{H}_{int}$. 

The tight-binding Hamiltonian on kagome lattice (see  \cref{lattice}) is given by
\begin{equation}\label{tight}
\begin{aligned}
\mathcal{H}_{0}=&-\mu\sum_{m}(a_{\mathbf{r}_{m}}^{(a)\dagger}a_{\mathbf{r}_{m}}^{(a)}+a_{\mathbf{r}_{m}}^{(b)\dagger}a_{\mathbf{r}_{m}}^{(b)}+a_{\mathbf{r}_{m}}^{(c)\dagger}a_{\mathbf{r}_{m}}^{(c)})\\
&-t\sum_{<m,n>}e^{i\varphi}(a_{\mathbf{r}_m}^{(b)\dagger}a_{\mathbf{r}_n}^{(a)}+a_{\mathbf{{r}_m}}^{(a)\dagger}a_{\mathbf{r}_n}^{(c)}+a_{\mathbf{{r}_m}}^{(c)\dagger}a_{\mathbf{{r}_n}}^{(b)})+h.c.
\end{aligned}
\end{equation}
Here $a_{\mathbf{r}_{m}}^{(\alpha)\dagger}$ and $a_{\mathbf{r}_{m}}^{(\alpha)}$($\alpha=a,b,c$) are creation and annihilation operators of a spinless fermion residing on sublattice $\alpha$ of the optical lattice and positioned at $\mathbf{r}_{m}$\cite{blakie2004adiabatic,blakie2005adiabatic,kohl2005fermionic}. $\mu$ is the chemical potential of loaded fermions, $t$ is the hopping parameter, and $\varphi$ is the phase of the hopping that can be tuned with the help of artificial gauge fields\cite{an2018engineering,lin2009synthetic,struck2012tunable,aidelsburger2011experimental}. The momentum space representation of the hopping Hamiltonian, $\mathcal{H}_{0}$, reads
\begin{equation}\label{Hk}
\begin{aligned}
&\mathcal{H}_{0}(\mathbf{k})
=-\mu(a_{\mathbf{k}}^{(a)\dagger}a_{\mathbf{k}}^{(a)}+a_{\mathbf{k}}^{(b)\dagger}a_{\mathbf{k}}^{(b)}+a_{\mathbf{k}}^{(c)\dagger}a_{\mathbf{k}}^{(c)})\\
&-2te^{i\varphi}\cos(\mathbf{k}\cdot\mathbf{b})a_{\mathbf{k}}^{(b)\dagger}a_{\mathbf{k}}^{(a)}-2te^{i\varphi}\cos(\mathbf{k}\cdot\mathbf{c})a_{\mathbf{k}}^{(a)\dagger}a_{\mathbf{k}}^{(c)}\\
&-2te^{i\varphi}\cos(\mathbf{k}\cdot\mathbf{c}-\mathbf{k}\cdot\mathbf{b})a_{\mathbf{k}}^{(c)\dagger}a_{\mathbf{k}}^{(b)}+h.c.\\
\end{aligned}
\end{equation}
Here $a_{\mathbf{k}}^{(\alpha)\dagger}$ and $a_{\mathbf{k}}^{(\alpha)}$($\alpha=a,b,c$) are creation and annihilation operators of a fermion on sublattice $\alpha$ with momentum $\mathbf{k}$, $\mathbf{b}=(\frac{1}{4},\frac{\sqrt{3}}{4})$, $\mathbf{c}=(\frac{1}{2},0)$, and $\mathbf{k}=(k_{x},k_{y})$.
The spectrum of the hopping Hamiltonian $\mathcal{H}_{0}$, which describes the lattice subject to the staggered flux depicted in \cref{lattice}, is found from the characteristic equation  $\det(E\mathbf{I}-\mathcal{H}_{0}(\mathbf{k}))=0$. The latter acquires the form
\begin{equation}
\begin{aligned}
x^{3}-\left(\frac{3}{2}+\frac{A(\mathbf{k})}{2}\right)x-\frac{1}{2}\left(\cos(3\varphi)+A(\mathbf{k})\cos(3\varphi)\right)=0,
\end{aligned}
\end{equation}
where $x=-\frac{E+\mu}{2t}$ and $A(\mathbf{k})=\cos k_{x}+2\cos\frac{k_{x}}{2}\cos\frac{\sqrt{3}k_{y}}{2}$.
If one of three energy bands of the Hamiltonian is non-dispersive (flat), the corresponding $x$ should be independent of $k$.  This implies $x+\cos(3\varphi)=0$,
$x^{3}-\frac{3}{2}x-\frac{1}{2}\cos(3\varphi)=0.$

Noting that $\varphi$ and $\varphi+2\pi/3$ are equivalent since $2\pi/3$ can be gauged out, and $\varphi$ and $-\varphi$ are related by time reversal transformation, the possible values for $\varphi$ supporting a flat band are $\varphi=0,\frac{\pi}{6},\pi$, and the corresponding band structures are shown in \cref{Pi0,Pi6,Pi1} where flat bands are located at top, middle, and bottom respectively. For $\varphi=0$ or $\pi$, the full Hamiltonian respects the time-reversal symmetry, while $\varphi=\pi/6$ breaks it, which leads to real or complex $J_{ijkl}$s as we will show later. At $\varphi=0$ or $\pi$ with an open boundary, a bandgap always exists because of the trivial topology\cite{bergman2008band}.

The real space expression of the impurity Hamiltonian corresponding to randomly  distributed    impurities reads
\begin{equation}\label{impurity}
\mathcal{H}_{imp}=u\sum_{{r}_{m}\in R}a_{\mathbf{r}_{m}}^{\dagger}a_{\mathbf{r}_{m}},
\end{equation}
where $u$ is the on-site potential and $R$ is a random set of $M$ sites with $M$ much smaller than $1/3$ of the number of the lattice sites, $\frac{\mathcal{L}}{3}$. In the following, we choose $M\sim0.1\mathcal{L}$.

The flat band of the kagome lattice originates from the structural destructive interference containing degenerate localized states circulating among the hexagons of the kagome lattice\cite{bergman2008band,MS}. The impurity Hamiltonian, $\mathcal{H}_{imp}$, in turn,  connects wave-functions between hexagons that are next to each other while keeping the localization property intact in general. Hence $\mathcal{H}_{imp}$ will remove a number of states from the flat band, but most of them will still remain there(\cref{spectrum}). For detailed calculations, see \cref{gap}. With rotational symmetry, the spectrum of $\mathcal{H}_{0}$ is sixfold. The impurity Hamiltonian $\mathcal{H}_{imp}$ breaks the rotational symmetry making the spectrum smeared. The impurity also widens the gap at the $\varphi=\pi$ case (see \cref{gap}) so that when considering the interaction one can project it onto the energy degenerate subset of the single-particle states at low energy.

Finally, the interaction Hamiltonian is given by 
\begin{equation}\label{interaction}
\mathcal{H}_{int}=\frac{1}{2} V \sum_{<m,n>}a_{\mathbf{r}_{m}}^{\dagger} a_{\mathbf{r}_{n}}^{\dagger} a_{\mathbf{r}_{m}} a_{\mathbf{r}_{n}},
\end{equation}
where the summation is over the nearest neighbors. Microscopically, the potential originates from the two-body scattering pseudo-potential as described in \cref{pseudo}. For potentials that decay fast enough, the scattering length determines the low energy scattering and the details of the interaction are irrelevant\cite{dalibard1999collisional,muller2011microscopic}.

\section{Emergent SYK physics}\label{SYK}

\begin{figure}[!ht]
\begin{subfigure}{0.23\textwidth}
\includegraphics[trim={5 10 5 10},clip,width=\textwidth]{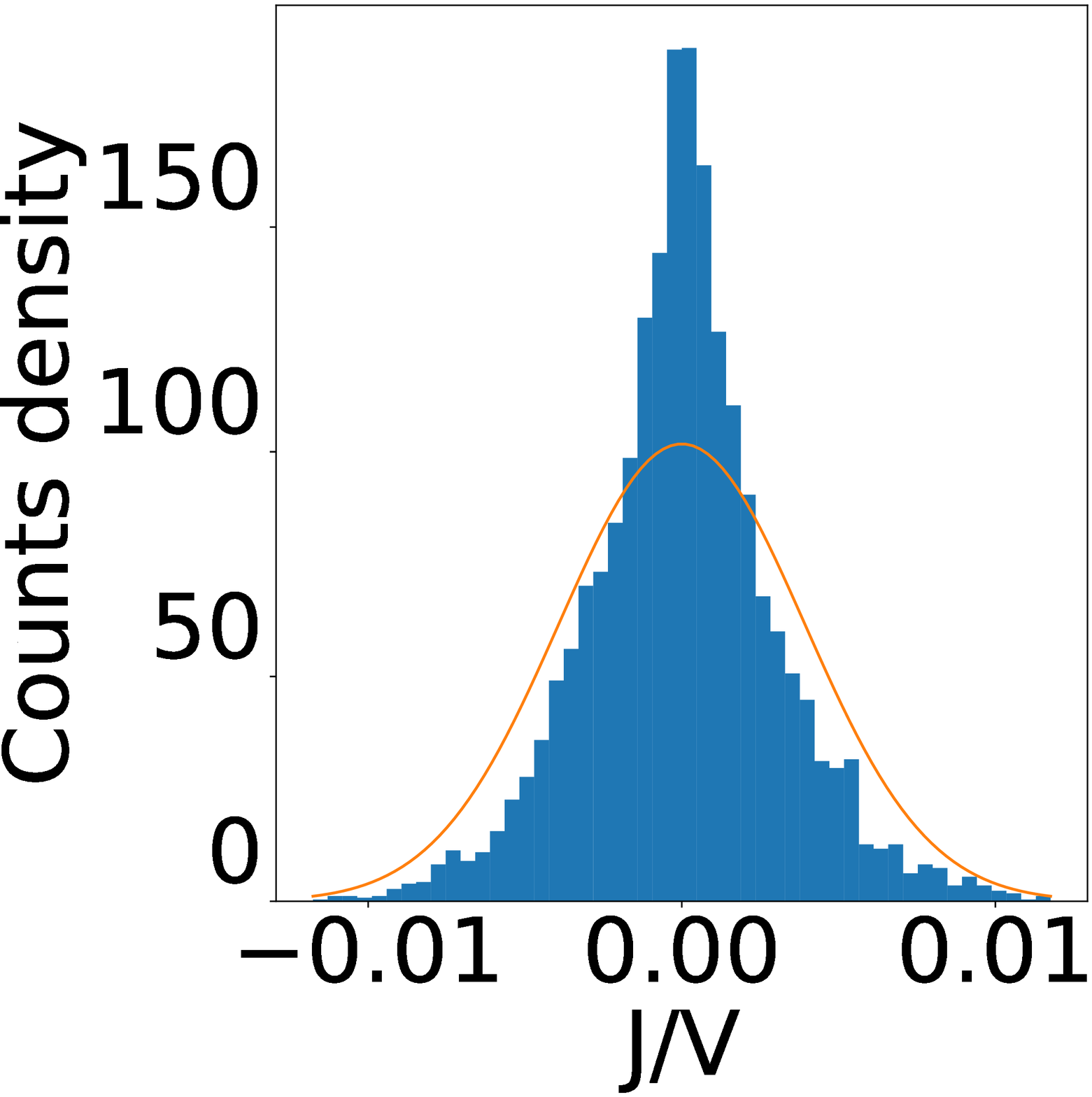}
\begin{picture}(0,0)
\put(-18,103){$(a)$}
\end{picture}
\phantomsubcaption{\label{J0}}
\vspace{-3mm}
\end{subfigure}
\begin{subfigure}{0.23\textwidth}
\includegraphics[trim={5 10 5 10},clip,width=\textwidth]{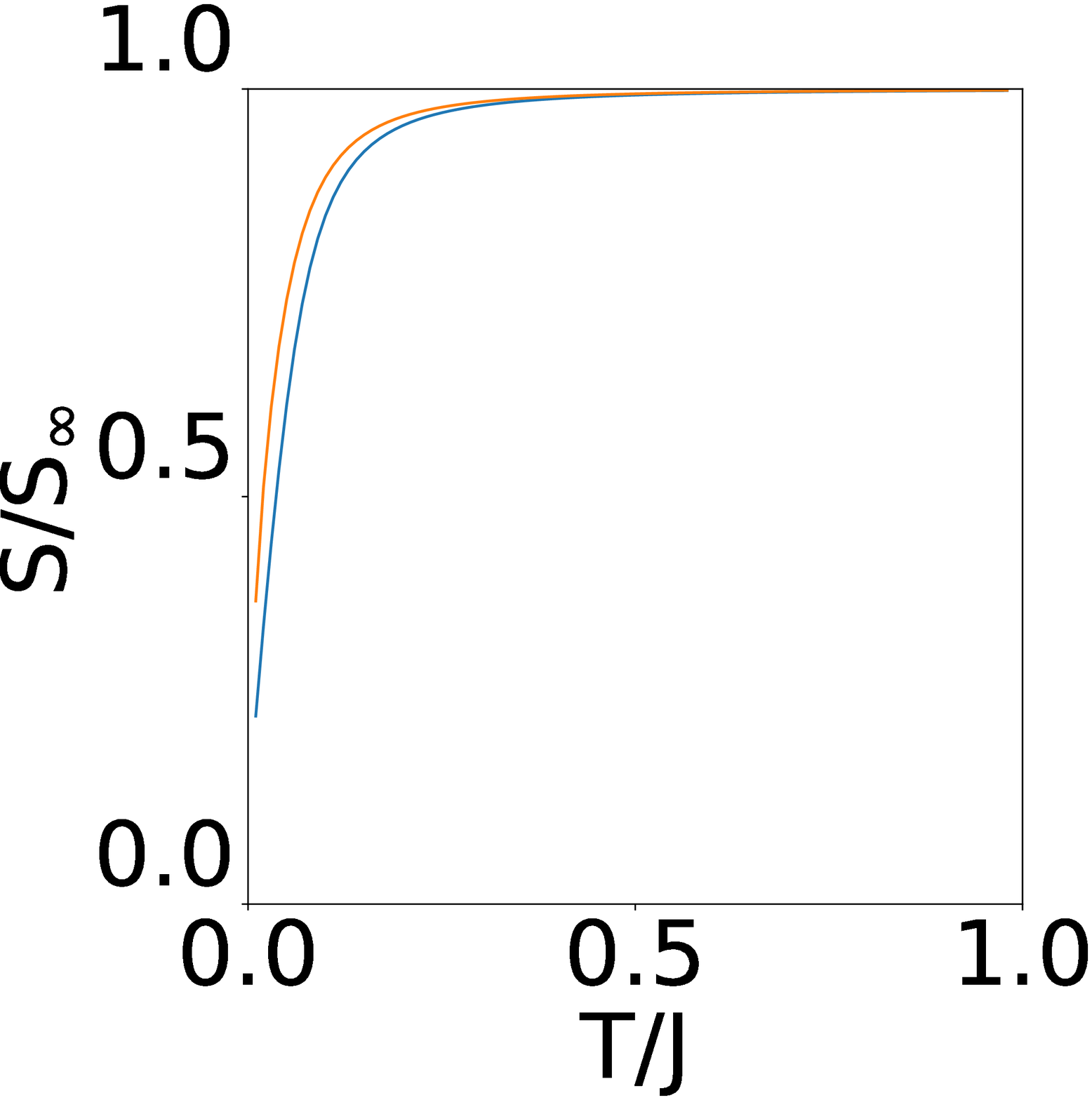}
\begin{picture}(0,0)
\put(35,100){$(b)$}
\end{picture}
\phantomsubcaption{\label{S0}}
\vspace{-3mm}
\end{subfigure}
\begin{subfigure}{0.23\textwidth}
\includegraphics[trim={5 10 5 10},clip,width=\textwidth]{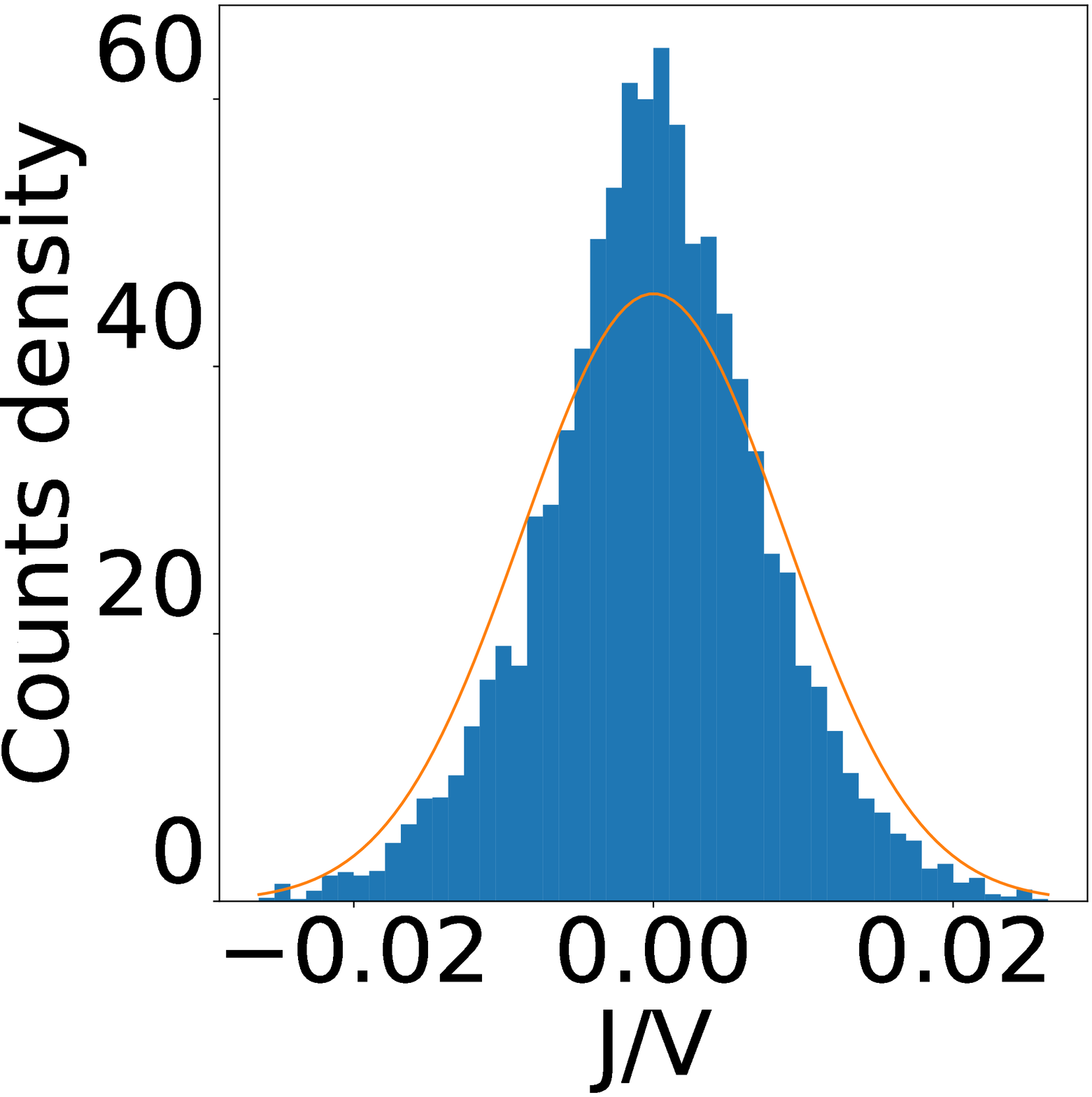}
\begin{picture}(0,0)
\put(-24,108){$(c)$}
\end{picture}
\phantomsubcaption{\label{J1}}
\vspace{-3mm}
\end{subfigure}
\begin{subfigure}{0.23\textwidth}
\includegraphics[trim={5 10 5 10},clip,width=\textwidth]{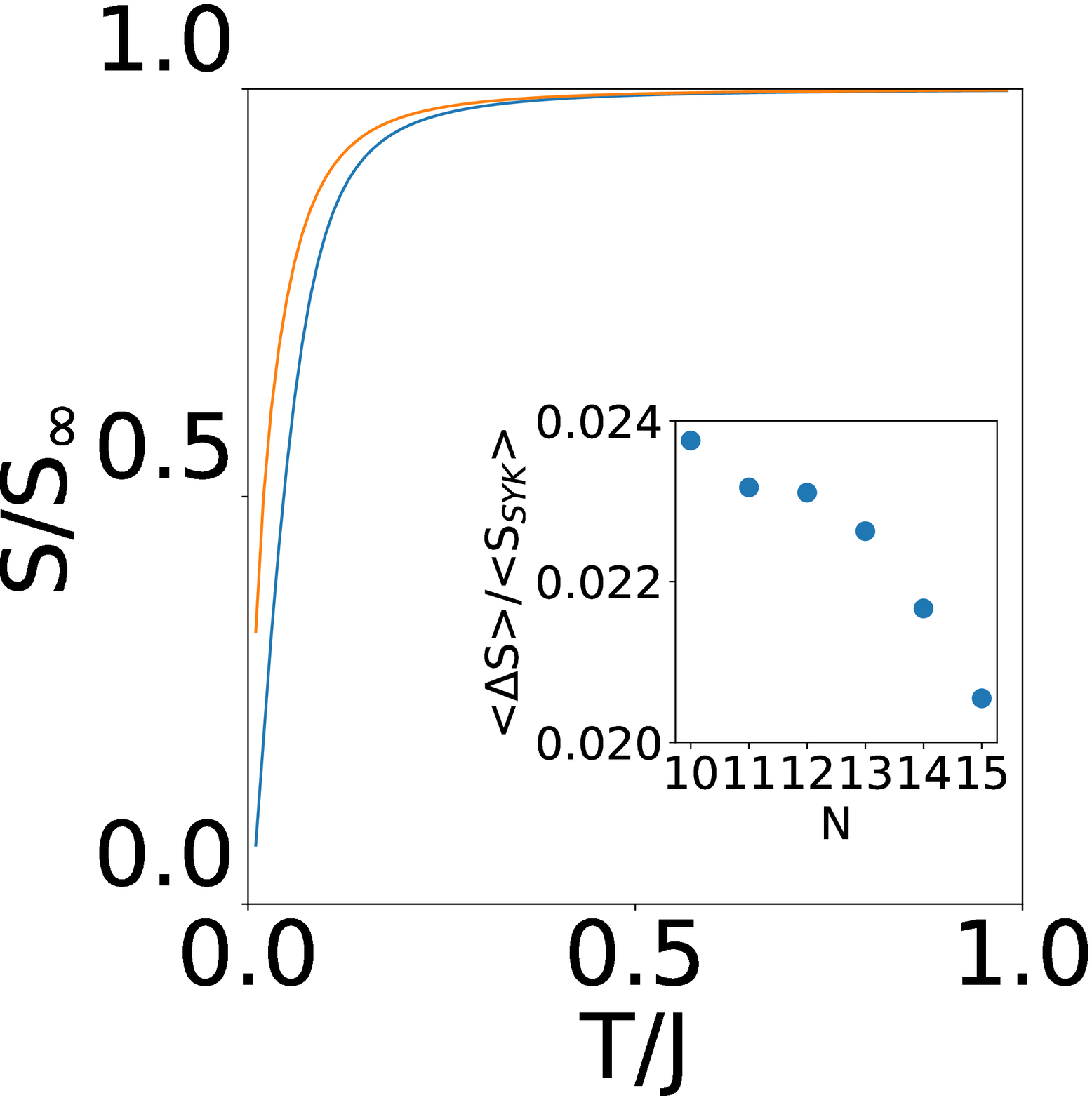}
\begin{picture}(0,0)
\put(34,100){$(d)$}
\end{picture}
\phantomsubcaption{\label{S1}}
\vspace{-3mm}
\end{subfigure}
\begin{subfigure}{0.23\textwidth}
\includegraphics[trim={0 10 0 10},clip,width=\textwidth]{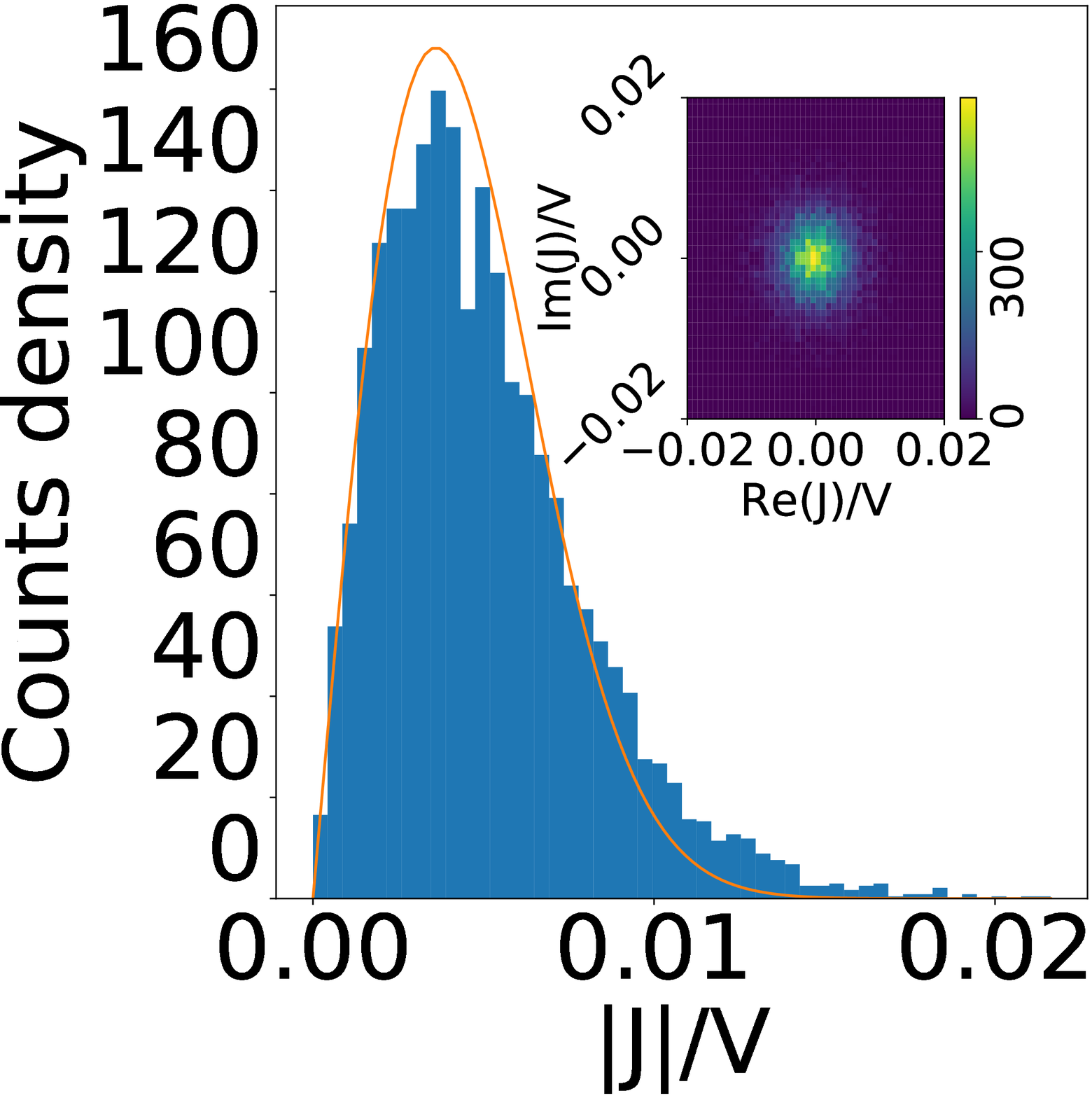}
\begin{picture}(0,0)
\put(-20,108){$(e)$}
\end{picture}
\phantomsubcaption{\label{J6}}
\end{subfigure}
\vspace{-2mm}
\begin{subfigure}{0.23\textwidth}
\includegraphics[trim={5 10 5 40},clip,width=\textwidth]{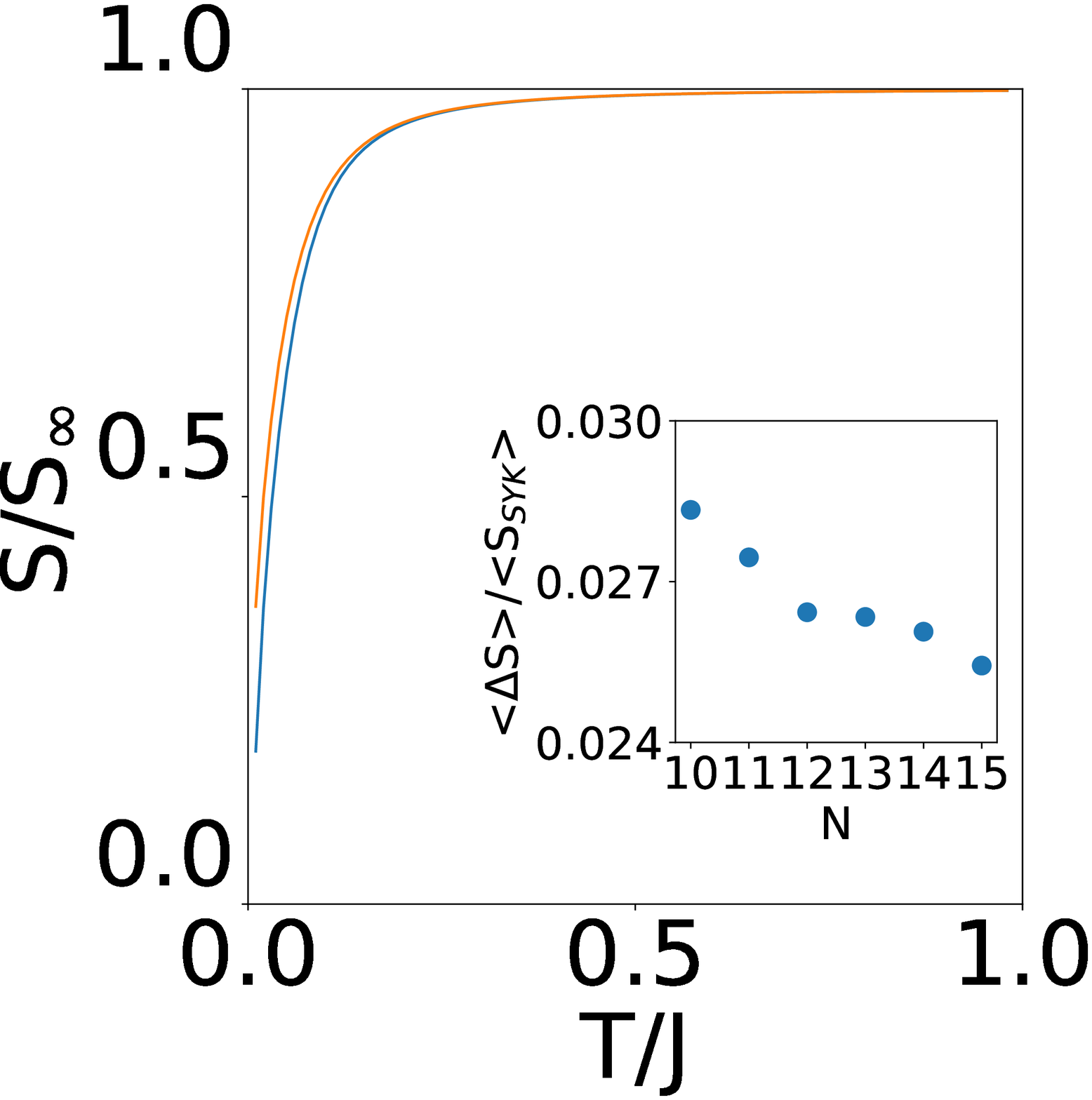}
\begin{picture}(0,0)
\put(32,100){$(f)$}
\end{picture}
\phantomsubcaption{\label{S6}}
\end{subfigure}
\vspace{-2mm}
\caption{(Color online). The results of the proposed optical lattice simulation are compared with the exact diagonalization results of the SYK model. Distribution of couplings  $J_{ijkl}$ (panels (a), (c), and (e)) and entropies (panels (b), (d), and (f)) are plotted for the proposed scheme described by the effective Hamiltonian \cref{eff}).
$J_{ijkl}s$ are measured in the units of $V$, which is the interaction strength of nearest neighbors.
For simulation the following parameters are chosen: $\varphi=0$ (panels (a) and (b)), $\varphi=\pi$ (panels (c)and (d)), and $\varphi=\pi/6$ (panels (e) and (f)), with $u=t$ and the number of states in flat band $N=15$. Inset to panel (e) shows the real and imaginary parts of couplings which are independent of each other and each of them is Gaussian distributed. In panels (b), (d), and (f) the upper (orange) lines correspond to the SYK model, and bottom (blue) lines are calculated from the effective Hamiltonian \cref{eff}. Insets to panels (d) and (f) show the decreasing averaged relative difference of entropy with respect to increasing $N$.
\label{JS}}
\end{figure}


\begin{figure}
\begin{subfigure}{0.49\textwidth}
\includegraphics[trim={5 10 5 10},clip,width=\textwidth]{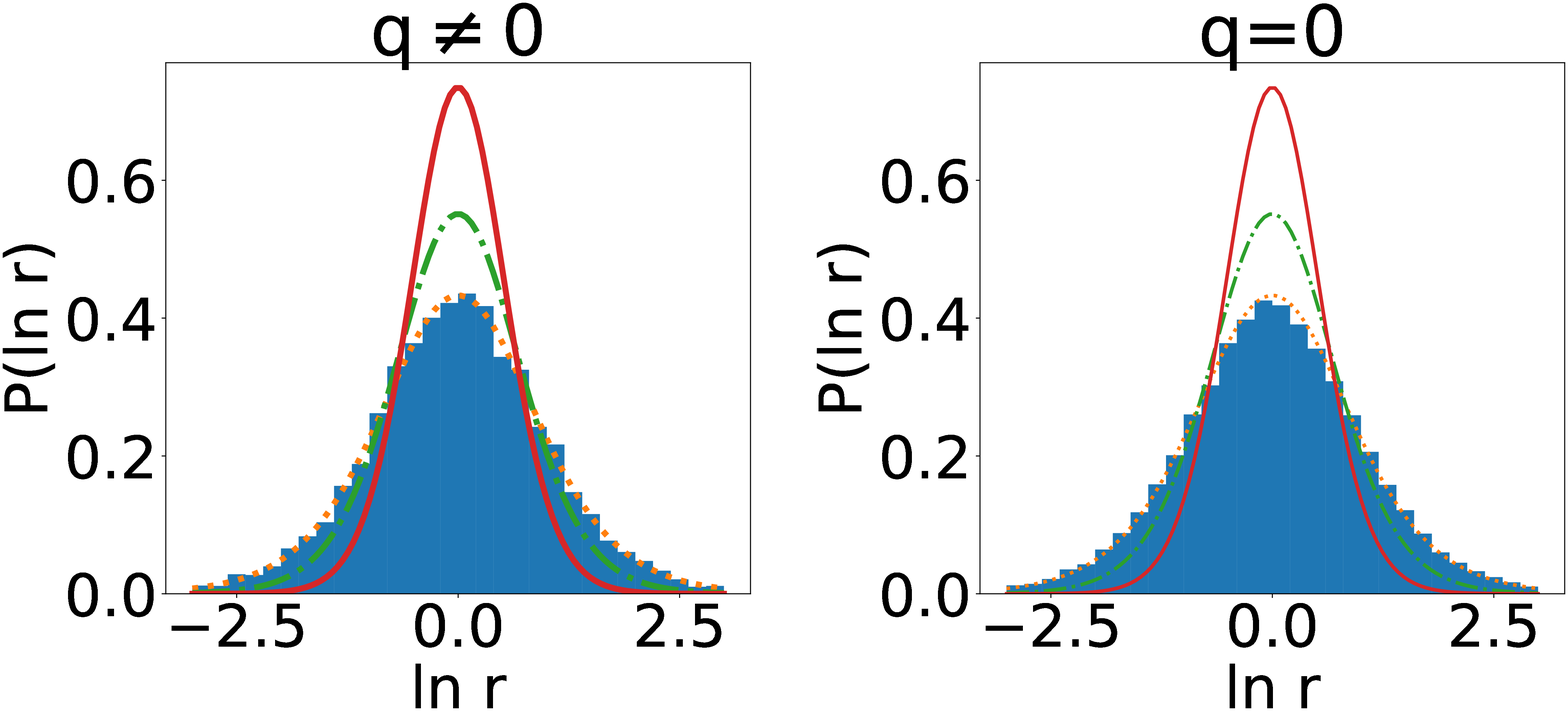}
\begin{picture}(0,0)
\put(-115,118){$(a)$}
\end{picture}
\phantomsubcaption{\label{r0}}
\vspace{-3mm}
\end{subfigure}
\begin{subfigure}{0.49\textwidth}
\includegraphics[trim={5 10 5 10},clip,width=\textwidth]{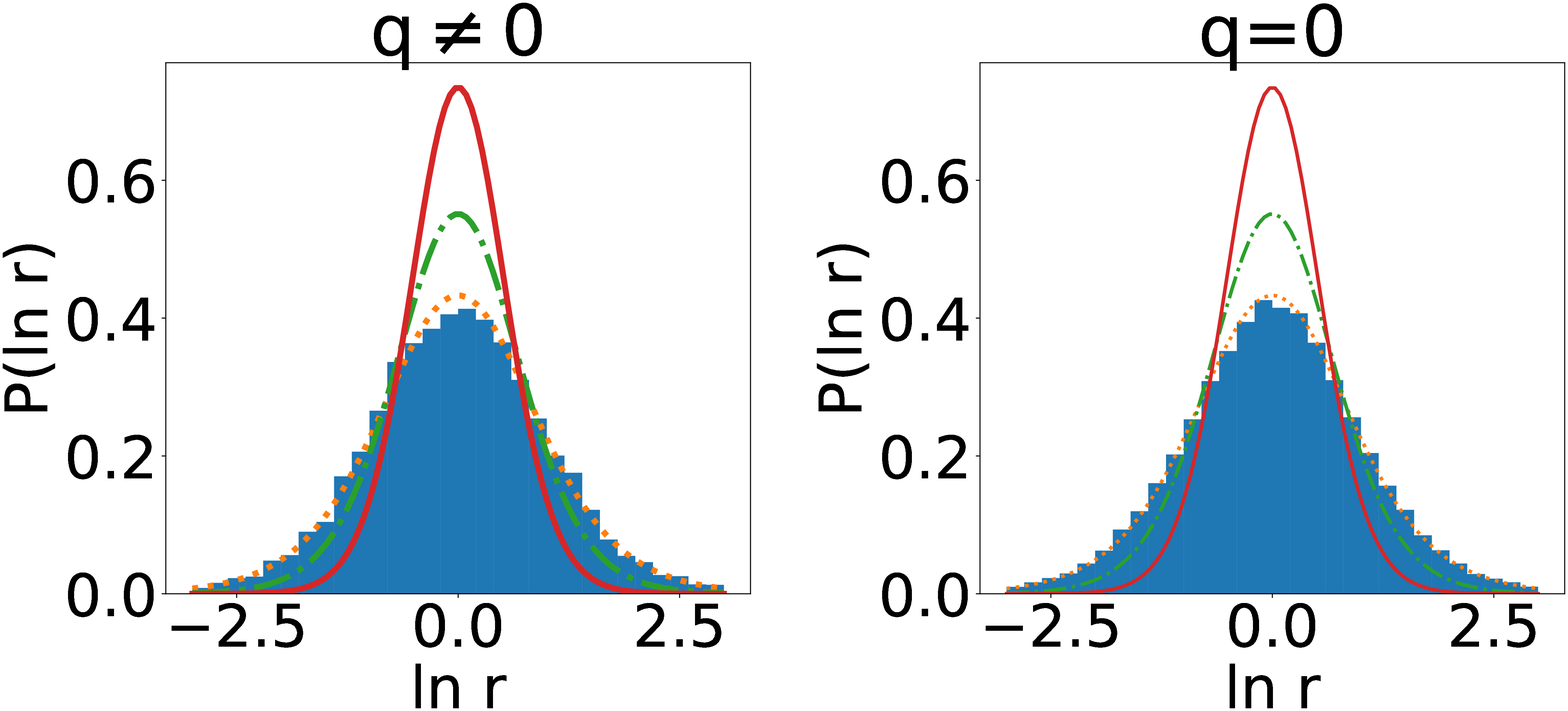}
\begin{picture}(0,0)
\put(-115,118){$(b)$}
\end{picture}
\phantomsubcaption{\label{r1}}
\vspace{-3mm}
\end{subfigure}
\begin{subfigure}{0.49\textwidth}
\includegraphics[trim={5 10 5 10},clip,width=\textwidth]{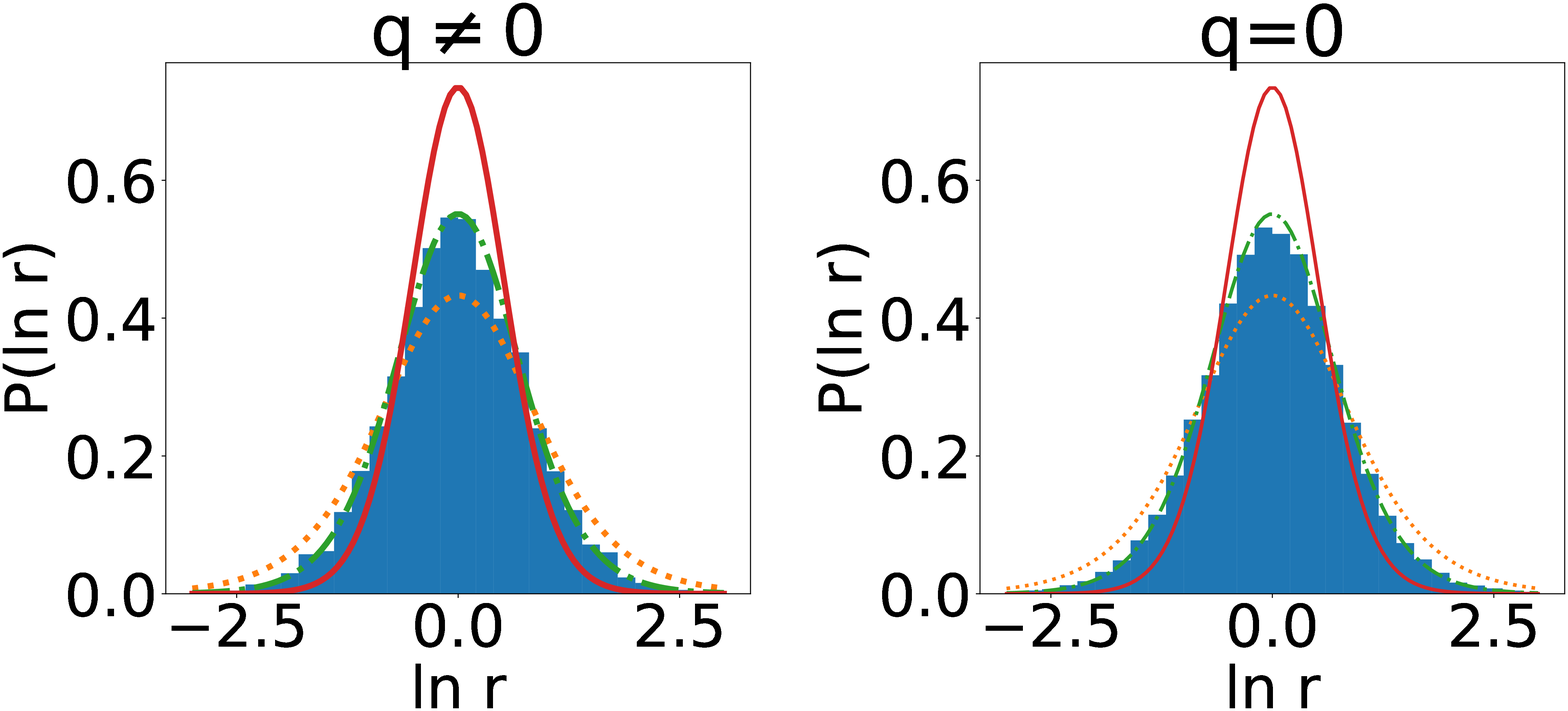}
\begin{picture}(0,0)
\put(-115,118){$(c)$}
\end{picture}
\phantomsubcaption{\label{r6}}
\end{subfigure}
\vspace{-2mm}
\caption{(Color online). Many-body level statistics shown as histograms for $N=14$, $u=t$, and $\varphi=0$ (panel (a)) , $\varphi=\pi$ (panel (b)), and $\varphi=\pi/6$ (panel (c)). Solid, dash-dotted and dotted curves correspond to GSE, GUE and GOE respectively.\label{r14}}
\end{figure}


\begin{figure}
\begin{subfigure}{0.23\textwidth}
\includegraphics[trim={5 10 5 10},clip,width=\textwidth]{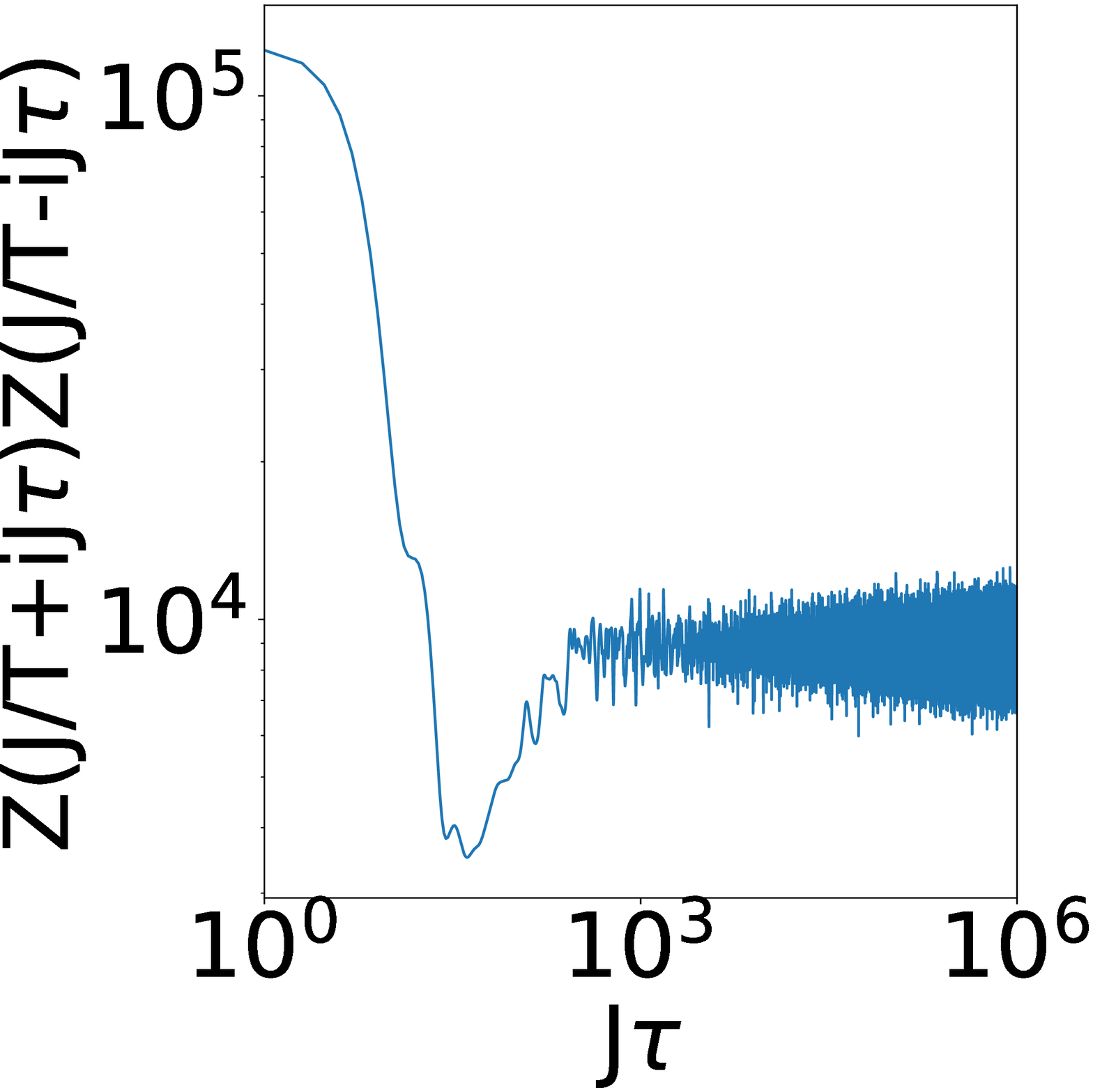}
\begin{picture}(0,0)
\put(34,110){$(a)$}
\end{picture}
\phantomsubcaption{}
\vspace{-3mm}
\end{subfigure}
\begin{subfigure}{0.23\textwidth}
\includegraphics[trim={5 10 5 10},clip,width=\textwidth]{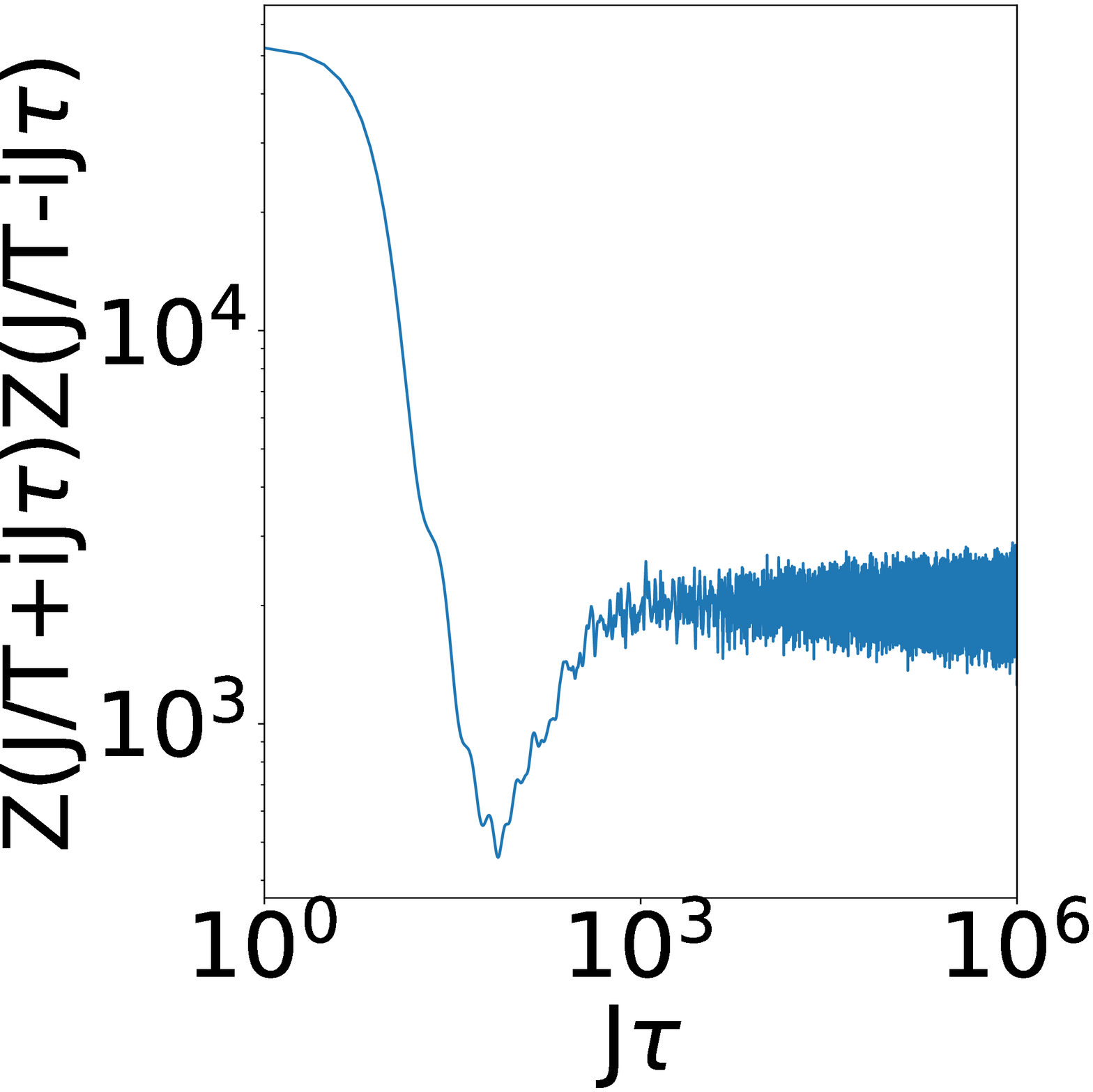}
\begin{picture}(0,0)
\put(34,110){$(b)$}
\end{picture}
\phantomsubcaption{}
\vspace{-3mm}
\end{subfigure}
\begin{subfigure}{0.23\textwidth}
\includegraphics[trim={5 10 5 10},clip,width=\textwidth]{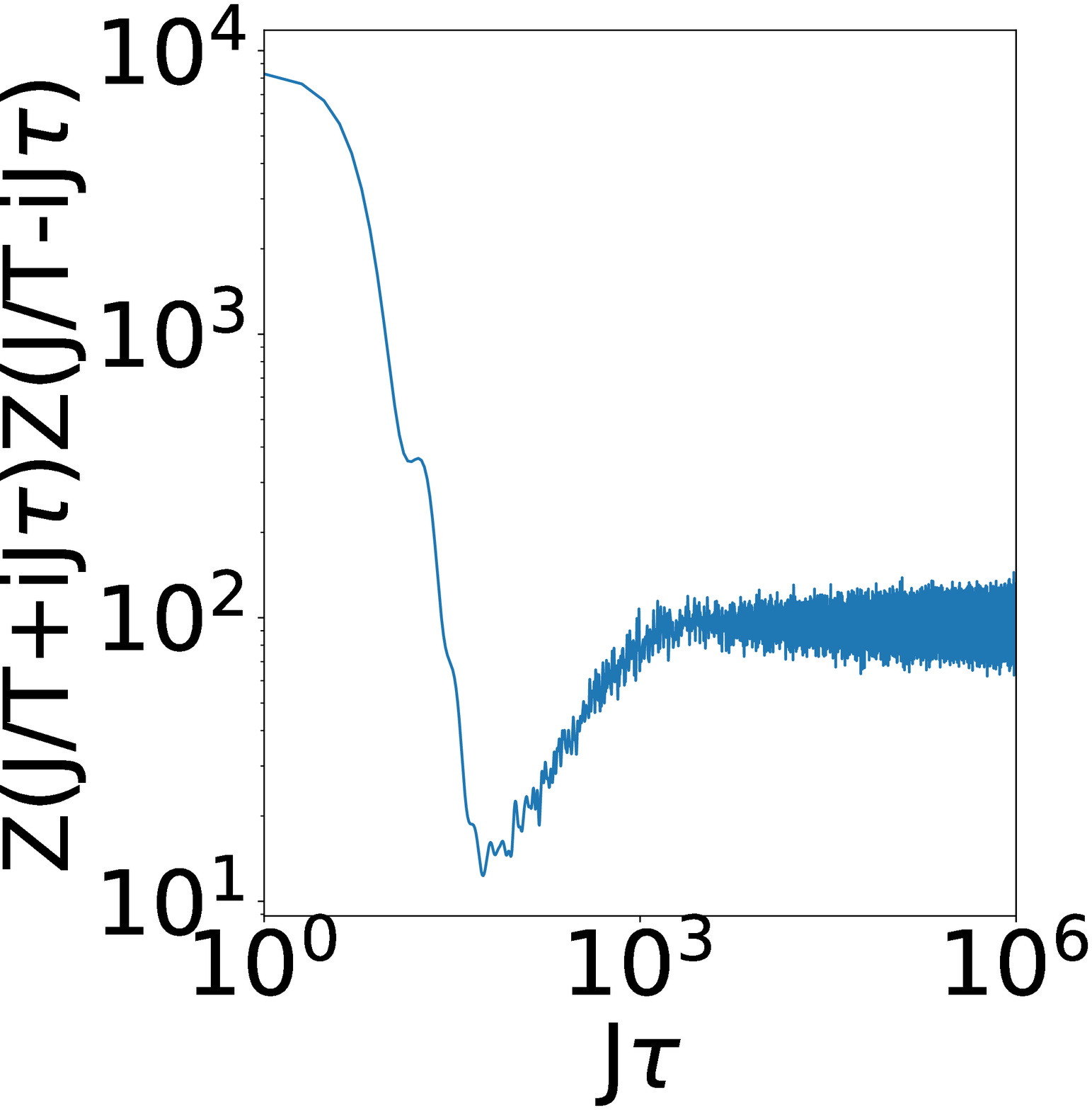}
\begin{picture}(0,0)
\put(34,110){$(c)$}
\end{picture}
\phantomsubcaption{}
\vspace{-3mm}
\end{subfigure}
\begin{subfigure}{0.23\textwidth}
\includegraphics[trim={5 10 5 10},clip,width=\textwidth]{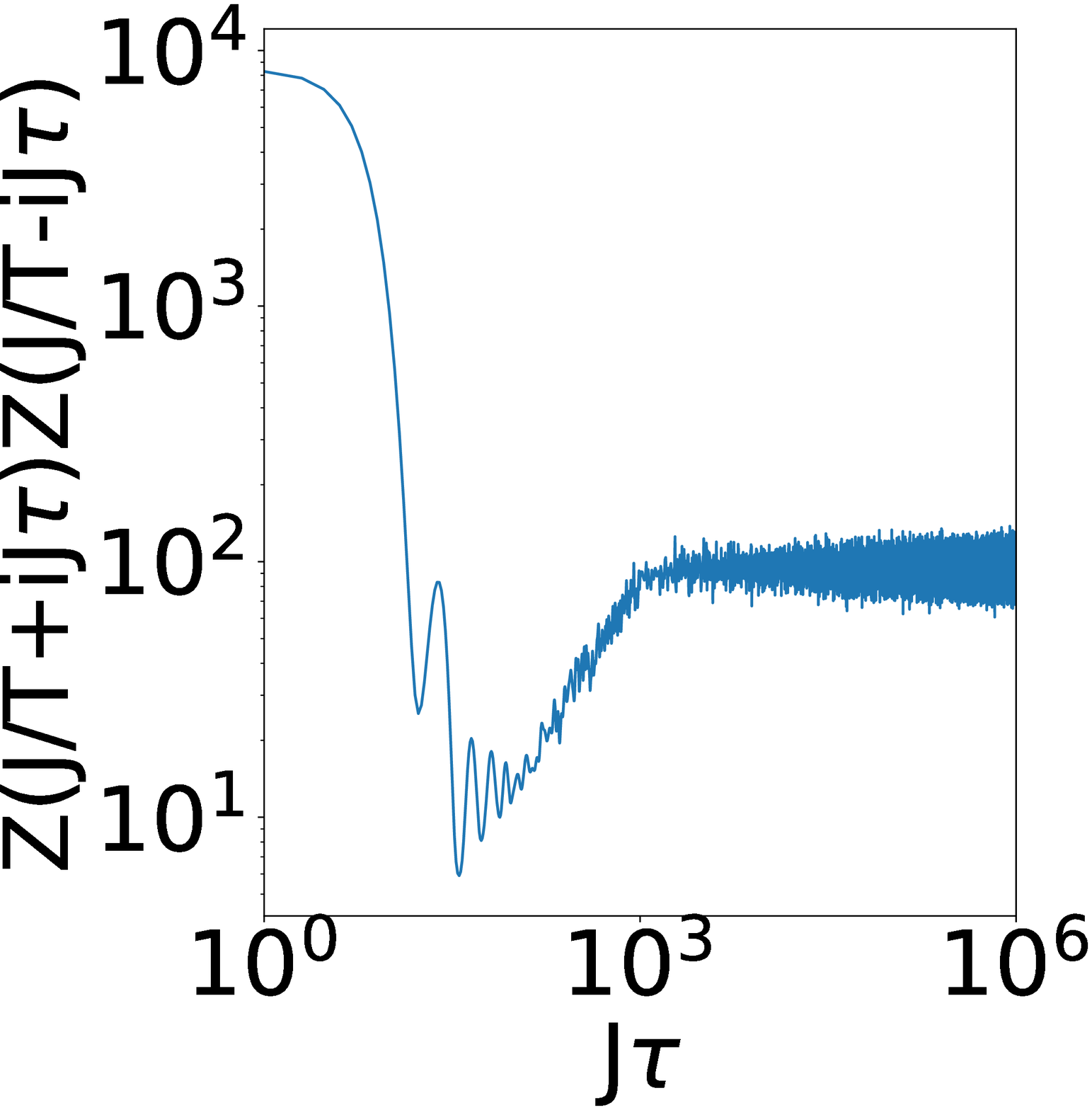}
\begin{picture}(0,0)
\put(34,110){$(d)$}
\end{picture}
\phantomsubcaption{}
\vspace{-3mm}
\end{subfigure}
\vspace{-2mm}
\caption{The spectral form factor is plotted at $\varphi=\pi$ (left column) and $\varphi=\pi/6$ (right column) for $J/T=10$ (inset (a) and (b)) and $J/T=1$ (inset (c) and (d)) for the system with the number of states in flat band $N=14$.\label{Z2}}
\end{figure}

\begin{figure}
\begin{subfigure}{0.23\textwidth}
\includegraphics[trim={5 10 5 10},clip,width=\textwidth]{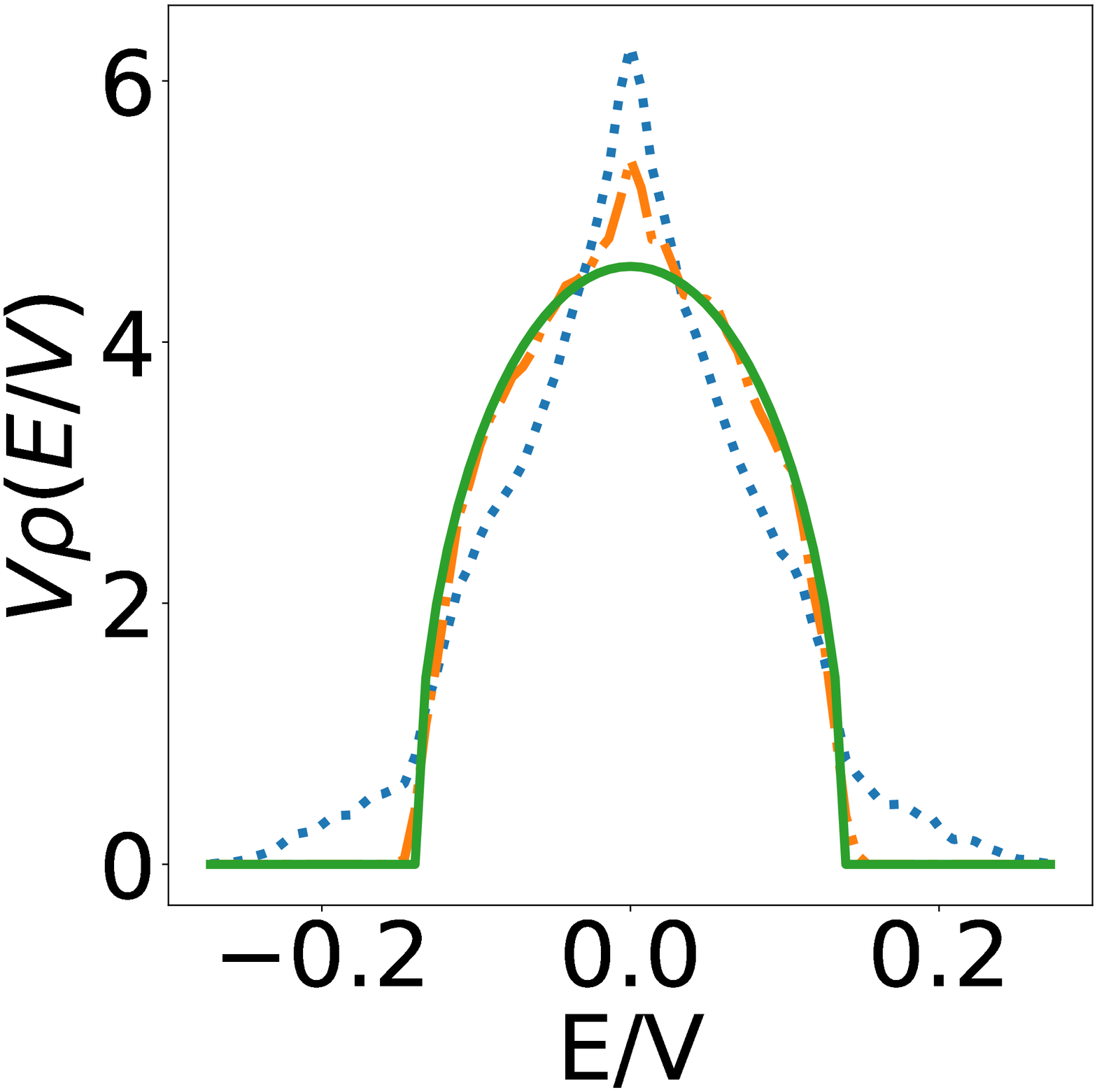}
\begin{picture}(0,0)
\put(-33,108){$(a)$}
\end{picture}
\phantomsubcaption{}
\end{subfigure}
\begin{subfigure}{0.23\textwidth}
\includegraphics[trim={5 10 5 10},clip,width=\textwidth]{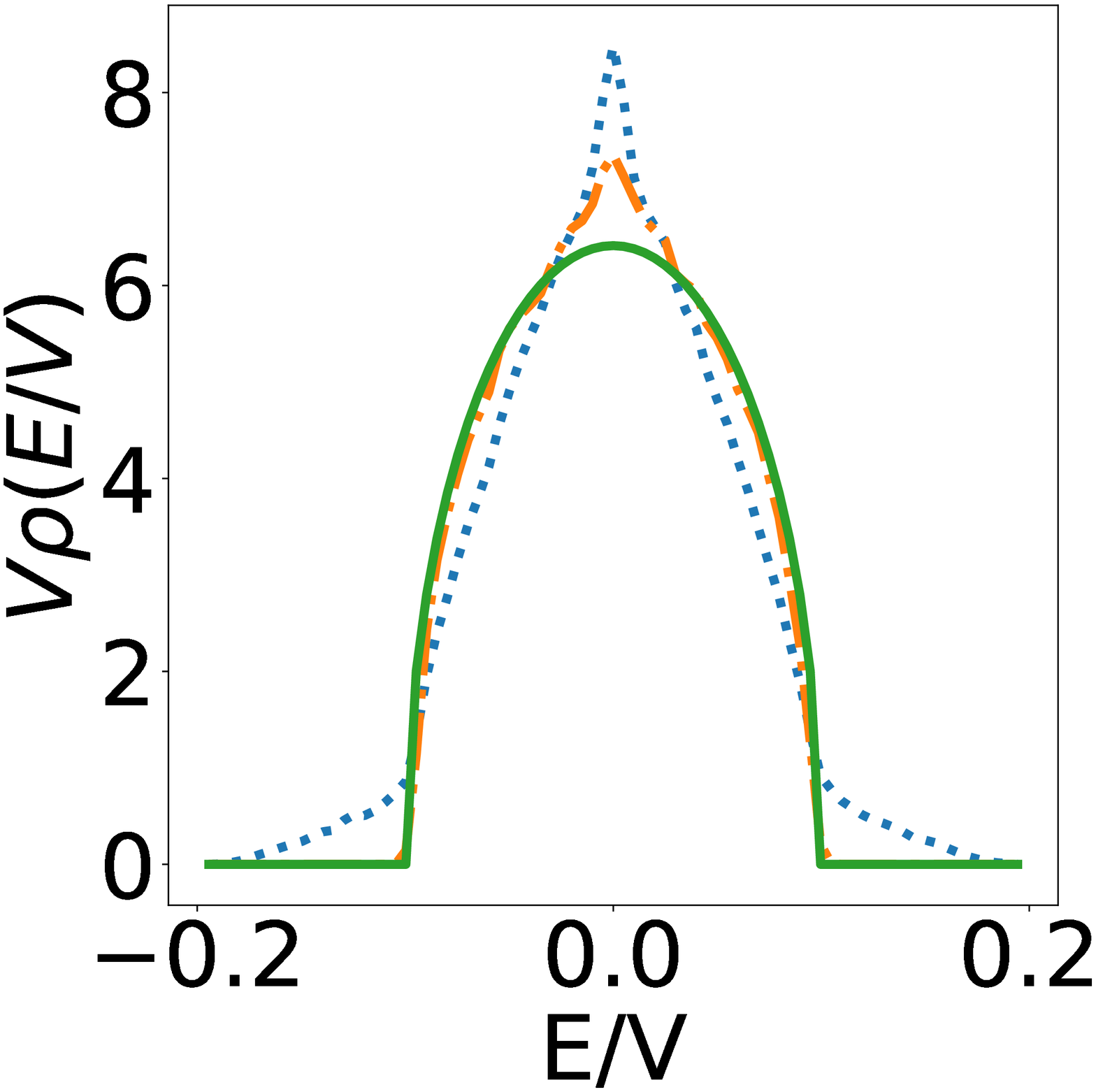}
\begin{picture}(0,0)
\put(-33,108){$(b)$}
\end{picture}
\phantomsubcaption{}
\end{subfigure}
\begin{subfigure}{0.23\textwidth}
\includegraphics[trim={5 10 5 10},clip,width=\textwidth]{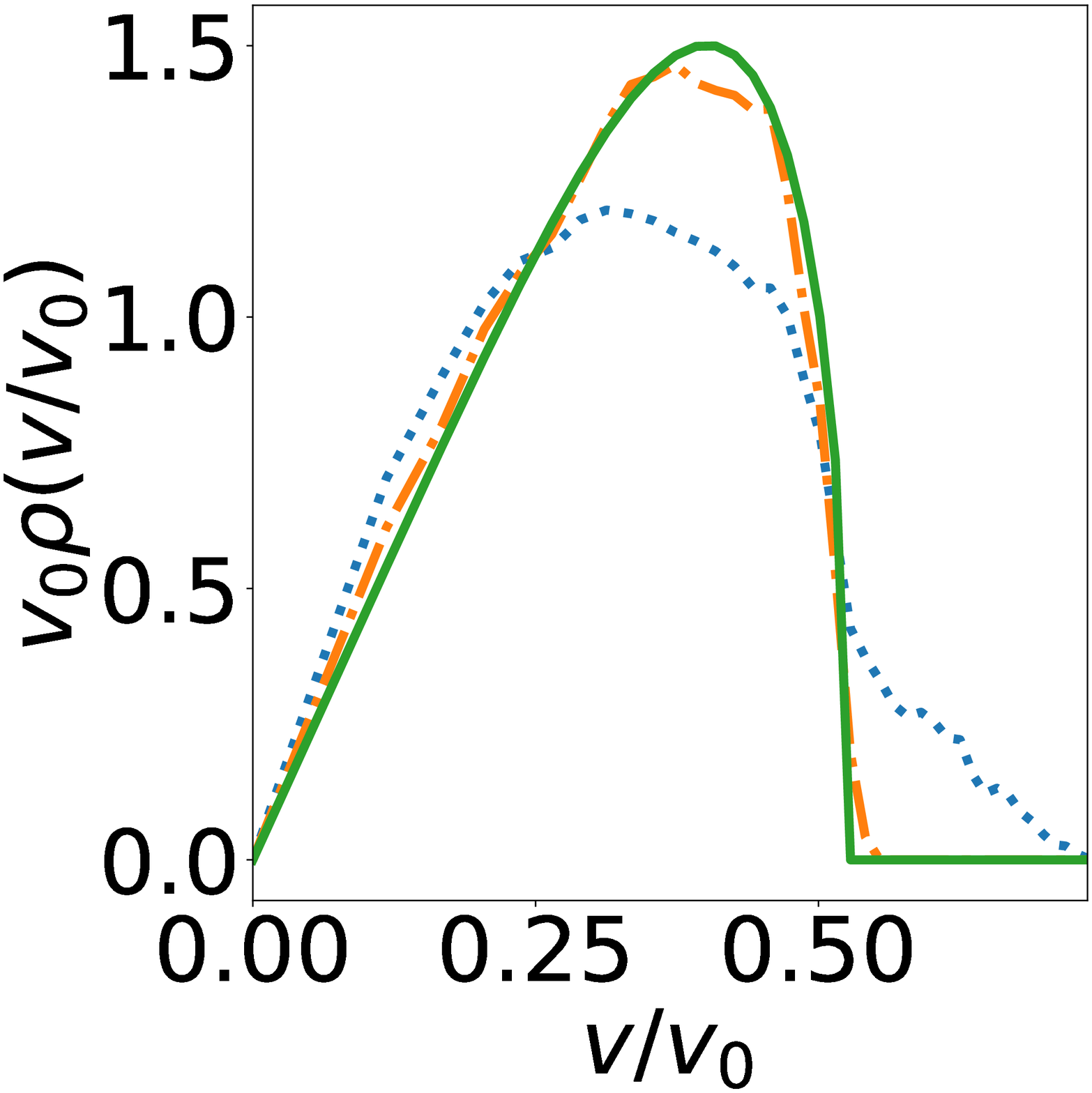}
\begin{picture}(0,0)
\put(-25,108){$(c)$}
\end{picture}
\phantomsubcaption{}
\end{subfigure}
\vspace{-2mm}
\begin{subfigure}{0.23\textwidth}
\includegraphics[trim={5 10 5 10},clip,width=\textwidth]{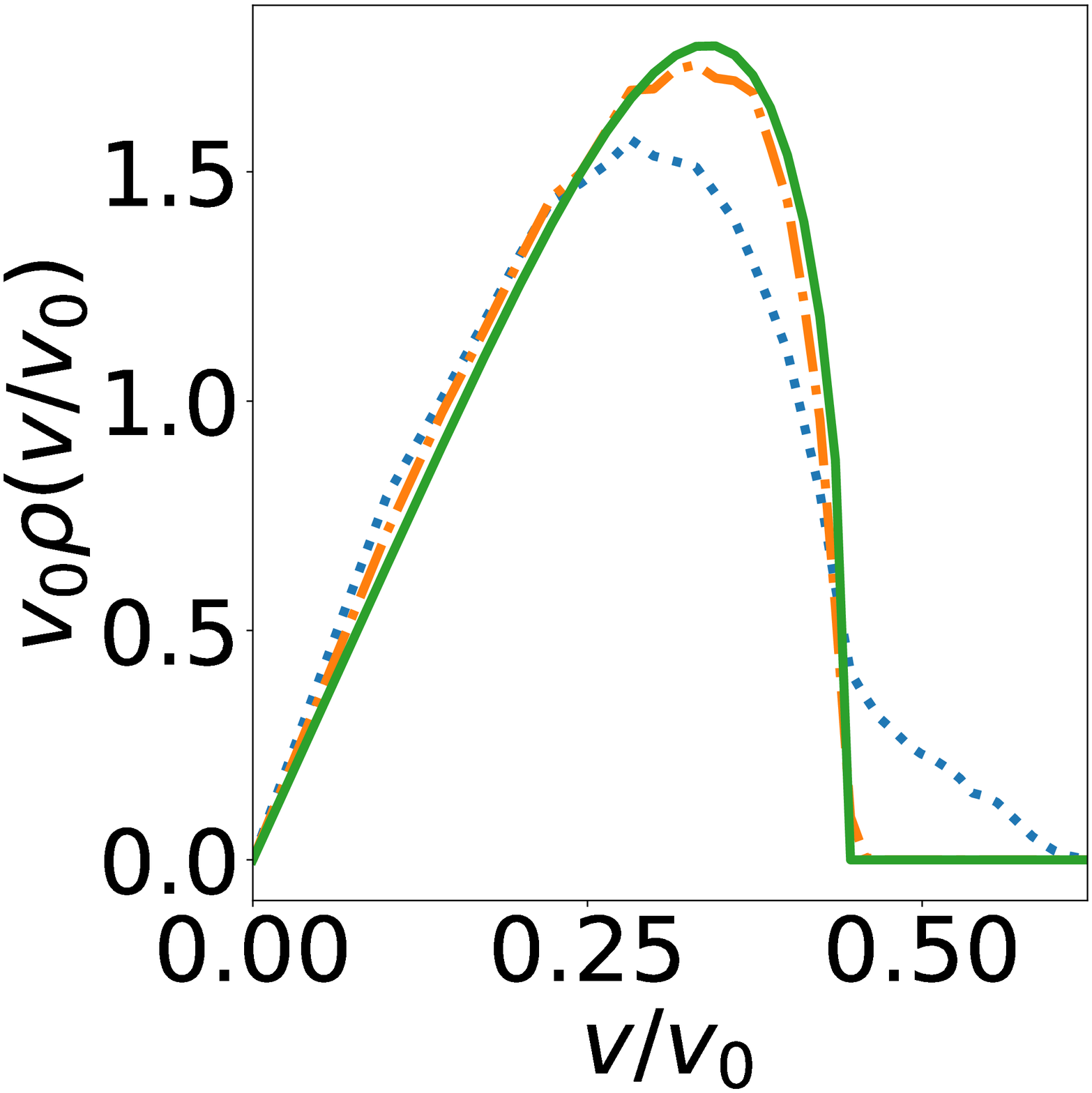}
\begin{picture}(0,0)
\put(-25,108){$(d)$}
\end{picture}
\phantomsubcaption{}
\end{subfigure}
\begin{subfigure}{0.23\textwidth}
\includegraphics[trim={5 10 5 10},clip,width=\textwidth]{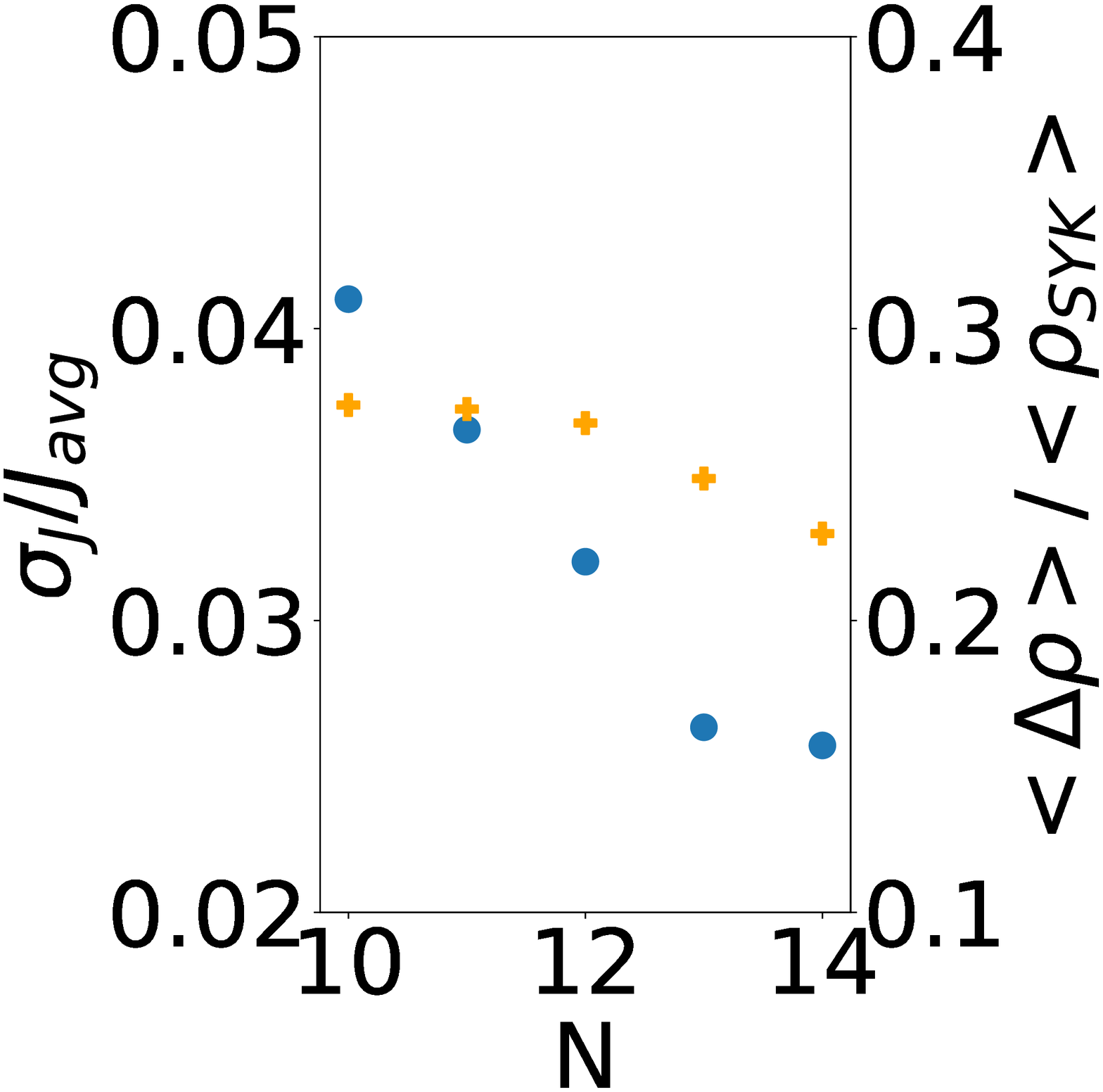}
\begin{picture}(0,0)
\put(-20,108){$(e)$}
\end{picture}
\phantomsubcaption{}
\end{subfigure}
\vspace{-2mm}
\begin{subfigure}{0.23\textwidth}
\includegraphics[trim={5 10 5 10},clip,width=\textwidth]{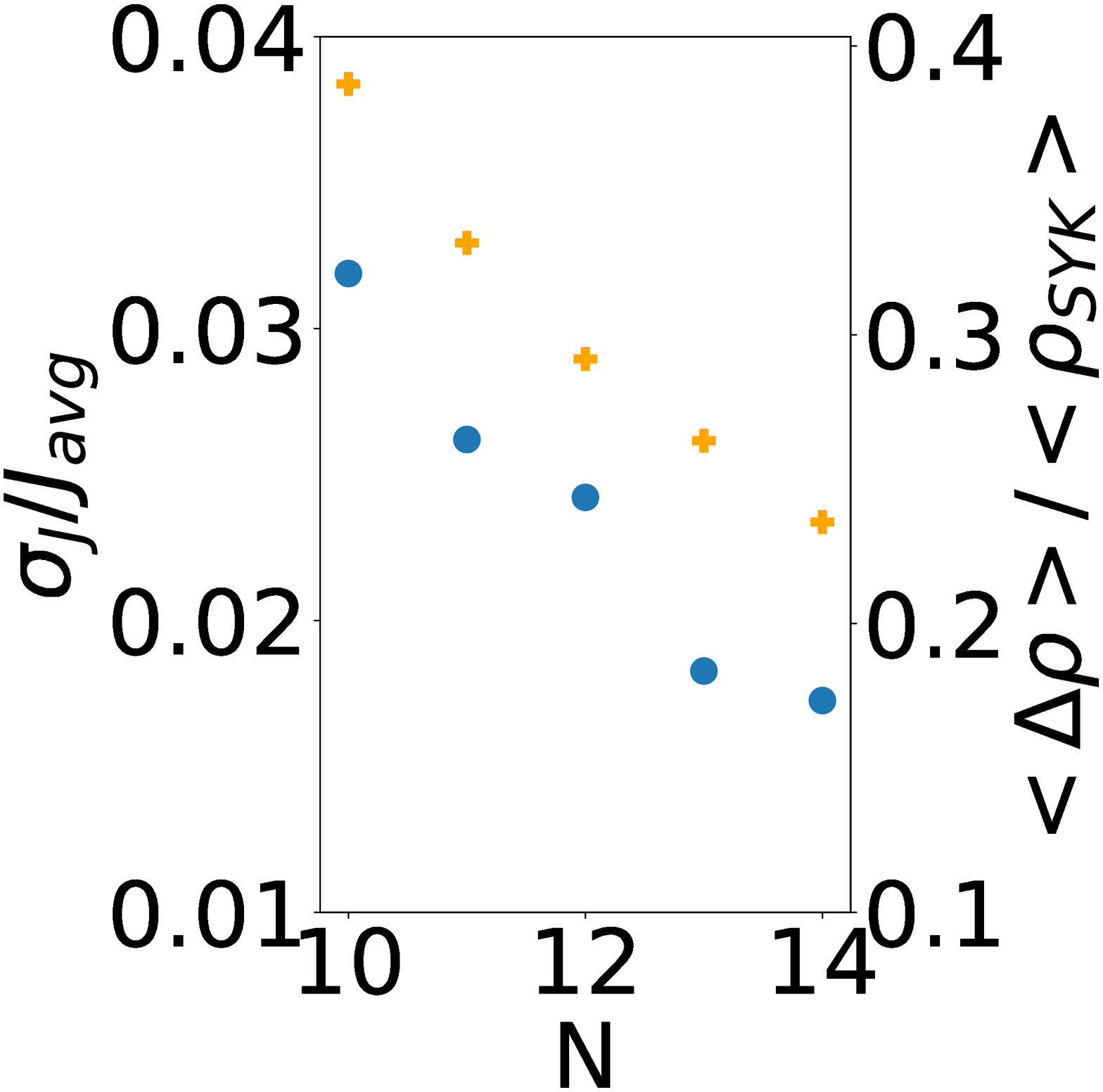}
\begin{picture}(0,0)
\put(-20,108){$(f)$}
\end{picture}
\phantomsubcaption{}
\end{subfigure}
\vspace{-2mm}
\caption{Energy spectral density (panels (a) and (b)) and the velocity distribution (panels (c) and (d)) after releasing the atoms (dotted curves correspond to the numerical results of the effective Hamiltonian \cref{eff}, dash-dotted curves correspond to the exact diagonalization results of the SYK model, and solid curves correspond to the Wigner semicircle law of the random matrix theory) at $N=14$. Panels (e) and (f) show the variance of $J$ (blue round dots) and the averaged relative difference of spectral density (orange crosses) decreases when N increases. Phases are $\varphi=\pi$ for panels (a),(c), and (e), and $\varphi=\pi/6$ for panels (b), (d), and (f). The velocity is measured in the units of $v_{0}=\sqrt{\frac{2\epsilon}{m}}$, where $m$ is the mass of the itinerant atoms.\label{Dis}}
\end{figure}

Now we will show that the free Hamiltonian, $\mathcal{H}_{0}+\mathcal{H}_{imp}$, together with the perturbation $\mathcal{H}_{int}$, is capable of generating the SYK model irrespective of the particular form of the interaction given that it is sufficiently weak.  Consider $N$ particles with wave-functions $\phi_{i}(\mathbf{r}_{m}), i=1,\cdots, N$,  that are accommodated within the flat band due to the fine-tuned chemical potential (e.g., keeping the lattice filling fraction $\nu\leq 1/3$ for  $\varphi=\pi$; $1/3< \nu\leq 2/3$ for $\varphi=\pi/6$; and $2/3<\nu\leq 1$ for $\varphi=0$).  The second quantized wave function of the fermion at cite ${\bf r}_m$ can be expanded over the basis of flat-band wave functions as $a_{\mathbf{r}_{m}}=\sum_{i} \phi_{i}({\bf r}_m) c_i$, where the $c_i$ is an annihilation operator of that state. In terms of these operators we can get the low-temperature effective Hamiltonian for the degenerate ground  states\cite{chen2018quantum}:
\begin{equation}\label{effTilde}
\mathcal{H}_{\text{eff}}=(2\tilde{t}-\mu)\sum_{i}c_{i}^{\dagger}c_{i}+\sum_{ijkl}\tilde{J}_{ijkl}c_{i}^{\dagger}c_{j}^{\dagger}c_{k}c_{l},
\end{equation}
with
\begin{equation}\label{JTilde}
\tilde{J}_{ijkl}=\frac{1}{2}V\sum_{<\mathbf{r}_{1},\mathbf{r}_{2}>}[\phi_{i}(\mathbf{r}_{1})\phi_{j}(\mathbf{r}_{2})]^{*}[\phi_{k}(\mathbf{r}_{1})\phi_{l}(\mathbf{r}_{2})],
\end{equation}
where $\tilde{t}=t$ for $\varphi=0$, $\tilde{t}=0$ for $\varphi=\pi/6$, and $\tilde{t}=-t$ for $\varphi=\pi$, $\phi_{i}(r)$ is the wave function of the $i$-th degenerate state, $r_{1/2}$ are the lattice sites, and $c_{i}^{\dagger}$ and $c_{i}$ are creation and annihilation operators of fermionic modes residing in the flat band.
Using the anti-commutation relations of creation and annihilation operators, it is convenient to equivalently rewrite the effective Hamiltonian $\mathcal{H}_{\text{eff}}$ as
\begin{equation}\label{eff}
\mathcal{H}_{\text{eff}}=(2\tilde{t}-\mu)\sum_{i}c_{i}^{\dagger}c_{i}+\sum_{i>j,k>l}J_{ijkl}c_{i}^{\dagger}c_{j}^{\dagger}c_{k}c_{l},
\end{equation}
where
\begin{equation}\label{J}
J_{ijkl}=\tilde{J}_{ijkl}+\tilde{J}_{jilk}-\tilde{J}_{jikl}-\tilde{J}_{ijlk}.
\end{equation}
Here  we introduced ordering of indices in the Hamiltonian, and now will show that the resultant couplings, $J_{ijkl}$, are fully random.
Suppose  $\tilde{J}_{ijkl}$ and $J_{ijkl}$ are calculated in the basis $\{\phi_{i}\}$. Under basis transformation $\phi_{i'}'=\sum_{i}U_{i'i}\phi_{i}$,
where $U_{i'i}$ is a unitary matrx, $U^{\dagger}U=1$, the couplings transform as
\begin{equation}
\tilde{J'}_{i'j'k'l'}=\sum_{ijkl}U_{i'i}^{*}U_{j'j}^{*}U_{k'k}U_{l'l}\tilde{J}_{ijkl},\nonumber
\end{equation}
\begin{equation}
J'_{i'j'k'l'}=\sum_{ijkl}U_{i'i}^{*}U_{j'j}^{*}U_{k'k}U_{l'l}J_{ijkl}.\nonumber
\end{equation}
We see that in the absence of impurities, the specific values of $J_{ijkl}$ are basis dependent. Fortunately, if couplings $J_{ijkl}$ are independent Gaussian random variables, $J'_{i'j'k'l'}$'s are also independent Gaussian random variables with the same variance. So the distribution will be independent of basis.

Interestingly, from \cref{J}, we have
\begin{equation}
\begin{aligned}
&\sum_{i>j,k>l}J_{ijkl}=\frac{1}{2}\sum_{<\mathbf{r}_{1},\mathbf{r}_{2}>}V\times\sum_{i>j,k>l}\\
&[\phi_{i}(\mathbf{r}_{1})\phi_{j}(\mathbf{r}_{2})-\phi_{j}(\mathbf{r}_{1})\phi_{i}(\mathbf{r}_{2})]^{*}[\phi_{k}(\mathbf{r}_{1})\phi_{l}(\mathbf{r}_{2})-\phi_{l}(\mathbf{r}_{1})\phi_{k}(\mathbf{r}_{2})]\\
&=\frac{1}{2}\sum_{<\mathbf{r}_{1},\mathbf{r}_{2}>}V|\sum_{i>j}(\phi_{i}(\mathbf{r}_{1})\phi_{j}(\mathbf{r}_{2})-\phi_{j}(\mathbf{r}_{1})\phi_{i}(\mathbf{r}_{2}))|^{2}.
\end{aligned}
\end{equation}

So $\sum_{i>j,k>l}J_{ijkl}$ and $V$ have the same sign and one may notice that the relation holds for a general potential also. This property implies that we have a constraint on the $J_{ijkl}$ couplings, which removes one degree of freedom. Fortunately, if the number of random $J_{ijkl}$ is large $(N\gg 1)$, the degree of freedom is large, and hence the constraint will not affect the statistical properties.

Consider now $n$ particles in the flat-band kagome lattice described by the Hamiltonian $\mathcal{H}_{0}+\mathcal{H}_{imp}$ with chemical potential $\mu$ that insures that the Fermi surface lies in the flat band. With given $n$, the conserved charge $Q=\sum_{i}(c_{i}^{\dagger}c_{i}-\frac{1}{2})$ has the eigenvalue $q=n-N/2$ ($N$ is the number of states in the flat band). For even $N$, $q$ would be an integer, and for odd $N$, $q$ would be a half-integer. Since $Q$ commutes with the Hamiltonian \cref{eff}, we can diagonalize the Hamiltonian within a specific $q$ subspace.

As the next step, we exactly diagonalize the Hamiltonian $\mathcal{H}_{0}+\mathcal{H}_{imp}$ with $q=-1/2$ or $0$ and calculate the couplings $J_{ijkl}$ using \cref{J} and three different interaction potentials. Then we calculate the thermodynamic entropy of the effective Hamiltonian \cref{eff}. The results for $\lambda=0.01a$ are shown in \cref{J0,S0} with phase $\varphi=0$ and \cref{J1,S1} with $\varphi=\pi$. In all cases, the distribution of $J_{ijkl}$ is nearly Gaussian, which is the defining property of the disordered couplings in the SYK model. In the SYK model, random couplings $J_{ijkl}$s are independent, and the distribution of a specific $J_{ijkl}$ will be the same as that of all the $J_{ijkl}$s in a single realization. Also, the entropy agrees with that of the SYK model obtained from exact diagonalization to some extent. One can further calculate the averaged relative difference of entropy $<\Delta S>/<S_{\mathrm{SYK}}>$ where $\Delta S=S_{\mathrm{SYK}}-S$ and $<S>=\int_{0}^{J} SdT/J$, as shown in the insets of \cref{S1,S6}. This is at the same order as that in Ref.~\onlinecite{chen2018quantum}, which is $0.018$ for $N=16$. One can expect that the difference will decrease as $N$ gets larger in the actual experimental setup. At the high-temperature limit $T/J\gg1$, the entropy approaches its maximum value  $S_{\infty}=k_B\ln\binom{N}{n}$, where $\binom{N}{n}$ is the binomial coefficient. As the temperature goes to zero, for infinite $N$, the entropy of the SYK model tends to a finite number. While for finite $N$, the entropy goes to zero as expected by the third law of thermodynamics. It can be shown that the averaged difference of entropies at finite $N$ and infinite $N$ is proportional to $1/N$\cite{fu2016numerical}.

The SYK Hamiltonian \cref{eff} without a chiral symmetry will experience no extra constraint when $q=0$\cite{you2017sachdev}. All the couplings $J_{ijkl}$ with chosen parameters can be real in a certain basis. So the probability distribution of energy spacings (defined as the distribution of $r_{n}=(E_{n+1}-E_{n})/(E_{n}-E_{n-1})$ where $E_{n}$ is the energy of the $n$th level) will exhibit behavior inherent to the Gaussian orthogonal ensemble (GOE)\cite{you2017sachdev}, as shown in the \cref{r0,r1}. For the $N=15$ case, $q$ is always non-zero. The level statistics is the same as for the $q\ne 0$ situation at $N=14$.

For $\varphi=\pi/6$, the same quantities, namely distribution of couplings $J_{ijkl}$, the entropy(\cref{J6,S6}), and level statistics(\cref{r6}) are calculated with $u=t$. Now one has complex couplings $J_{ijkl}$ whose real part and imaginary part are approximately Gaussian and the distribution of $|J_{ijkl}|$ becomes a $\chi$ distribution with degree of freedom two which indicates the independence of real and imaginary parts. Also the level statistics follows the distribution of the Gaussian unitary ensemble(GUE)\cite{you2017sachdev}.

While the distribution of $r_n$ shows the correlations between adjacent energy levels, the spectral form factor defined as
\begin{equation}
\begin{aligned}
&Z(J/T+iJ\tau)Z(J/T-iJ\tau)\\
=&\sum_{n, m} e^{-(J/T+iJ\tau) E_{n}} e^{-(J/T-iJ\tau) E_{m}},
\end{aligned}
\end{equation}
where $Z$ is the partition function of the model, characterizes correlations between all energy levels at all scales\cite{cotler2017black}. On the gravity side, this quantity describes the properties of the black hole in the dual AdS space\cite{papadodimas2015local,saad2018semiclassical}.

The averaged spectral form factor $\left\langle|Z(J/T+iJ\tau)|^{2}\right\rangle_{J}$ is shown in \cref{Z2}. One can see that at short times the slope regime is dominated by the decoupled SYK saddle points\cite{saad2018semiclassical}, for which it decays with a power law. The late time ramp and plateau originate from the statistics of the random matrix ensemble. Similar behavior exists in the Jackiw-Teitelboim gravity\cite{saad2018semiclassical}. This behavior is robust against temperatures ranging from $0.1J$ to $J$.

Importantly, the averaged two-point OTOC function, $\int dA\left\langle A(0)A^{\dagger}(t)\right\rangle_{J}$, where $A$ is a local unitary operator one can access in experiments and $dA$ is the Haar measurement with respect to it\cite{cotler2017chaos},
is proportional to the spectral form factor $|Z(2T,\tau)|^{2}$. OTOC can be measured by the many-body Loschmidt echo scheme\cite{swingle2016measuring,garttner2017measuring}.

Upon releasing the trap and the optical lattice in time-of-flight experiments, the fermion states are projected onto plane waves, and the many-body energy spectrum distribution of the flat-band system (given by the Wigner semicircle) is projected to the energy distribution of atoms.
Hence, the velocity distribution of atoms, as free particles, can be determined from many-body Wigner-Dyson statistics, as seen in \cref{Dis}. The latter can be observed in the time of flight experiments. We notice that the distribution of energy and the velocity predicted for the present scheme has a somewhat long tail, which is attributed to the small variation of $J$ when averaging over different disordered realizations. Our numerical simulations suggest that, when $N$ increases, the variation of $J$ is averaged out and the averaged relative difference of spectral density $<\Delta\rho>/<\rho_{\mathrm{SYK}}>=\int|\rho-\rho_{\mathrm{SYK}}|dE/\int\rho_{\mathrm{SYK}}dE$ decreases, leading to a faster decaying tail.

\section{Discussion of experimental aspects}{\label{experiment}}

In the following, we discuss aspects of the experimental realization of the proposed scheme:
\begin{enumerate*}[label=\roman*)]
\item The randomly distributed on-site potential can be realized by heavy atoms randomly loaded in the optical lattice whose strength can be tuned by Feshbach resonances\cite{hara2013three}.
\item Positive or complex hopping can be realized with the help of artificial gauge fields created by applying a zero averaged  homogeneous inertial force\cite{struck2012tunable} or Raman-assisted tunneling in asymmetric kagome lattice\cite{aidelsburger2011experimental}.
\item While each of the required steps namely, loading immobile impurities to random sites of the optical lattice, introducing the pseudo-gauge fields, and creating an optical lattice of kagome type are by now standard experimental techniques, combining all of these into one experiment may require some additional efforts.
\item To access the conformal limit in experiments, the parameter $J$ should be tuned through Feshbach resonances\cite{Tiecke2010broad, mckay2011cooling} that are large enough($J \gg k_BT$).  For fermionic atoms such as $^6$Li or $^{40}$K, and an experimentally accessible temperature of order $T \approx 1 nK$\cite{jo2012ultracold, weld2009spin, yang2020cooling}, one can estimate employing \cref{Vap} the required scattering length of order $a_p \approx 0.35a$.
\item One can, in principle, measure the observables when $\phi=0$ or $\phi=\pi/6$ by doubling the number of experimental runs. In the first run, the chemical potential should be kept slightly lower than the flat band, and, for the second one, the chemical potential should be kept in the flat band.
Upon analyzing data and assuming that the Fermi surface does not experience a phase transition, one should subtract the contribution of the Fermi sea in the measured observables to detect the emergent SYK physics. This would be difficult for the $\phi=0$ case since the system will have a small gap when $\mathcal{L}$ becomes large. So the $\phi=\pi/6$ and $\phi=\pi$ cases would be better choices for experiments, with the latter requiring less experimental runs.
\end{enumerate*}

The realization of the SYK model within the described technique is not specific to the kagome lattice. We believe these phenomena are quite universal, and the described technique can be applied to other lattices supporting a flat band. This, for example, can be seen upon investigating SYK physics from the interplay of disorder and interactions in the experimentally realized Lieb lattice using trapped fermions\cite{taie2015coherent}.  

\begin{acknowledgments}
The research was supported by startup funds from UMass Amherst. We are grateful to the anonymous reviewer of Physical Review A for their insightful comments. 
\end{acknowledgments}

\appendix
\section{Effect of an on-site impurity\label{gap}}

Hamiltonians contributed from the impurities on different sites commutate with each other, so one can separate their contribution within a perturbative scheme. So, consider the tight-binding Hamiltonian for $\varphi=\pi$ and one impurity on an arbitrary site: $\mathcal{H}=\mathcal{H}_{0}+\mathcal{H}_{imp}$:
\begin{equation}
\begin{aligned}
&\mathcal{H}_{0}=-\mu\sum_{m}a_{\mathbf{r}_m}^{\dagger}a_{\mathbf{r}_m}+t\sum_{<m,n>}a_{\mathbf{r}_m}^{\dagger}a_{\mathbf{r}_n},\\
&\mathcal{H}_{imp}=ua_{\mathbf{R}}^{\dagger}a_{\mathbf{R}},
\end{aligned}
\end{equation}
where ${\mathbf{R}}$ represents the location of the impurity.

The eigenstates of $\mathcal{H}_{0}$ in the flat band can be written as\cite{bergman2008band}
\begin{equation}
\ket{\mathbf{r}}=\frac{1}{\sqrt{6}}(\sum_{m}(-1)^{m}a_{\mathbf{r}_{m}}^{\dagger})\ket{0}.
\end{equation}
where $\mathbf{r}$ represents the center of the hexagonal plaquette of the kagome lattice, and $r_{m}$ goes over all six lattice cites around the hexagon\cite{bergman2008band}. Note that these localized states are not orthogonal, in fact
\begin{equation}
\bra{\mathbf{r}}\ket{\mathbf{r}'}=\delta_{\mathbf{r},\mathbf{r}'}-\frac{1}{6}\delta_{\mathbf{r},\mathbf{r}+\mathbf{\delta}}.
\end{equation}

The eigenstates of $\mathcal{H}_{0}$  in the dispersive band that touch the flat band can be written as
\begin{equation}
\ket{{\mathbf{k}}}=(\beta^{(a)}(\mathbf{k})a_{\mathbf{k}}^{(a)\dagger}+\beta^{(b)}(\mathbf{k})a_{\mathbf{k}}^{(b)\dagger}+\beta^{(c)}(\mathbf{k})a_{\mathbf{k}}^{(c)\dagger})\ket{0}.
\end{equation}
By employing the anti-commutation relation
\begin{equation}
\{a_{\mathbf{R}}^{(\alpha)},a_{\mathbf{k}}^{(\beta)\dagger}\}=\{a_{\mathbf{R}}^{(\alpha)},\sum_{\mathbf{r}}e^{i\mathbf{k}\mathbf{r}} a_{\mathbf{r}}^{(\beta)\dagger}\}=e^{i\mathbf{k}\mathbf{R}}\delta_{\alpha,\beta},
\end{equation}
the matrix elements of $\mathcal{H}_{imp}$ become
\begin{equation}
\bra{\mathbf{r}}ua_{\mathbf{R}}^{\dagger}a_{\mathbf{R}}\ket{\mathbf{r}'}=
\begin{cases}
0\text{ if }\mathbf{R}\notin\{\mathbf{r}_{m}\}\cap\{\mathbf{r}'_{m}\}\\
\frac{1}{3}u\delta_{\mathbf{r},\mathbf{r}'}-\frac{1}{6}u\text{ if }\mathbf{R}\in\{\mathbf{r}_{m}\}\cap\{\mathbf{r}'_{m}\}
\end{cases},
\end{equation}
\begin{equation}\label{decoupled}
\bra{\mathbf{k}}ua_{\mathbf{R}}^{\dagger}a_{\mathbf{R}}\ket{\mathbf{r}}=0,
\end{equation}
\begin{equation}
\bra{\mathbf{k}'}ua_{\mathbf{R}}^{\dagger}a_{\mathbf{R}}\ket{\mathbf{k}}=u\beta^{(\alpha)*}(\mathbf{k}')\beta^{(\alpha)}(\mathbf{k})e^{i(\mathbf{k}-\mathbf{k}')\mathbf{R}}.
\end{equation}

From \cref{decoupled}, one can see that the flat band and the dispersive one are decoupled. This means that the effect of impurity can be treated separately. The energy difference (compared to the unperturbed energy) $\Delta E$ of the flat band subset in the presence of the impurity can be found from
\begin{equation}
\det(\Delta E
\begin{pmatrix}
1 & -\frac{1}{6}\\
-\frac{1}{6} & 1
\end{pmatrix}
-u
\begin{pmatrix}
\frac{1}{6} & -\frac{1}{6}\\
-\frac{1}{6} & \frac{1}{6}
\end{pmatrix})=0,
\end{equation}
which leads to $\Delta E=0$ or $\frac{2}{7}u$. 
This analytical analysis explains slight lifting of the degeneracy of the flat-band states presented in numerical data of \cref{spectrum}.

The energy difference of the band touching  due to the impurity to the lowest order of $u$ is $\Delta E=u|\beta^{(\alpha)}(\mathbf{k})|^{2}$. As a special case, the gap around $\mathbf{k}=0$ would become $\Delta E=u|\beta^{(\alpha)}(0)|^{2}$.
In conclusion, the impurity will lift part of the flat band states and widen the gap, which is positive and proportional to $u$.

\section{Estimate for the interaction strength from a pseudo-potential\label{pseudo}}
If the optical lattice is loaded with spinless fermions, the potential energy coming from two-body scatterings can be written as
\begin{equation}
V = \frac{1}{2} \int d\mathbf{r}_1 d\mathbf{r}_2 \hat{\Psi}^{\dagger}(\mathbf{r}_1) \hat{\Psi}^{\dagger}(\mathbf{r}_2) \hat{V}_p(\mathbf{r}_1 - \mathbf{r}_2) \hat{\Psi}(\mathbf{r}_2) \hat{\Psi}(\mathbf{r}_1).
\end{equation}
Here $\hat{V}_p$ is the $p$-wave pseudo-potential\cite{idziaszek2006pseudopotential} given by
\begin{equation}
\hat{V}_p(\mathbf{r}) = \frac{2\pi \hbar^2 a_p^{3}}{m} \overleftarrow{\nabla} \delta^3(\mathbf{r}) \overrightarrow{\nabla} r \frac{\partial^3}{\partial r^3} r^2,
\end{equation}
which takes into account all the higher-order perturbations. The state
$\hat{\Psi}(\mathbf{r})$ can be expanded in the basis of flat-band states as
\begin{equation}
\hat{\Psi}(\mathbf{r}) = \sum_{i}\sum_{m} \phi_i(\mathbf{r}_m) W(\mathbf{r} - \mathbf{r}_m) c_{i},
\end{equation}
where $W(\mathbf{r} - \mathbf{r}_m)$ is the Wannier function for the lattice. One can easily check that the anti-commutation relation $\{ \hat{\Psi}(\mathbf{r}), \hat{\Psi}(\mathbf{r}^{\prime}) \} = \delta(\mathbf{r} - \mathbf{r}^{\prime})$ is satisfied with the help of orthonormality of $\phi_i(\mathbf{r}_m)$ and the $W(\mathbf{r} - \mathbf{r}_m)$ and anti-commutation relation of $c_i$.

With the above preparation, one is ready to get the Hamiltonian \ref{effTilde} with 
\begin{equation} \label{app.J}
\tilde{J}_{ijkl} = \frac{1}{2} \sum_{m \ne n, p \ne q} \phi_i^{*}(\mathbf{r}_m) \phi_j^{*}(\mathbf{r}_n) V_{dp}(\mathbf{r}_m, \mathbf{r}_n, \mathbf{r}_p, \mathbf{r}_q) \phi_k(\mathbf{r}_p) \phi_l(\mathbf{r}_q),
\end{equation}
where $m = n$ or $p = q$ terms vanish because of the Fermionic nature of the Hamiltonian, and the discrete pseudo-potential
\begin{equation}
\begin{aligned}
V_{dp}(\mathbf{r}_m, \mathbf{r}_n, \mathbf{r}_p, \mathbf{r}_q) =& \int d\mathbf{r}_1 d\mathbf{r}_2 W^{*}(\mathbf{r}_1 - \mathbf{r}_m) W^{*}(\mathbf{r}_2 - \mathbf{r}_n)\\
&\times \hat{V}_p(\mathbf{r}_1 - \mathbf{r}_2) W(\mathbf{r}_2 - \mathbf{r}_p) W(\mathbf{r}_1 - \mathbf{r}_q).
\end{aligned}
\end{equation}

For deep enough potential, the Wannier function is approximately Gaussian,
\begin{equation}
W(\mathbf{r}) = (\frac{m\omega}{\pi\hbar})^{\frac{1}{2}} (\frac{m\omega_z}{\pi\hbar})^{\frac{1}{4}} e^{-\frac{m\omega}{2\hbar} (x^2 + y^2) -\frac{m\omega_z}{2\hbar} z^2}.
\end{equation}
It is then straightforward to show that
\begin{equation}
\begin{aligned}
V_{dp}(\mathbf{r}_m, \mathbf{r}_n, \mathbf{r}_p, \mathbf{r}_q) = \frac{12\pi \hbar^2 a_p^3}{m} \int d\mathbf{r}_2 (\frac{m \omega_z}{\hbar})^2 z_2^2\\
\times W(\mathbf{r}_2 - \mathbf{r}_m) W(\mathbf{r}_2 - \mathbf{r}_n) W(\mathbf{r}_2 - \mathbf{r}_p) W(\mathbf{r}_2 - \mathbf{r}_q).
\end{aligned}
\end{equation}
Assuming the overlap between the localized states at different sites is small, the leading term of the discrete pseudo-potential
\begin{equation}
\begin{aligned}
V_{dp}(\mathbf{r}_m, \mathbf{r}_n, \mathbf{r}_p, \mathbf{r}_q) = \frac{12\pi \hbar^2 a_p^3}{m} \delta_{mq}\delta_{np}  \delta_{<mn>} \int d\mathbf{r}_2 (\frac{m \omega_z}{\hbar})^2 z_2^2\\
W(\mathbf{r}_2 - \mathbf{r}_m) W(\mathbf{r}_2 - \mathbf{r}_n) W(\mathbf{r}_2 - \mathbf{r}_n) W(\mathbf{r}_2 - \mathbf{r}_m),
\end{aligned}
\end{equation}
where the $\delta_{<mn>}=1$ only when $m$ and $n$ are indices corresponding to nearest neighbors while for all other cases $\delta_{<mn>}=0$.
Then one can introduce a new variable $\mathbf{d} = \mathbf{r}_m - \mathbf{r}_n$ and perform integration over $\mathbf{r}_2$,
\begin{equation}
\begin{aligned}
V_{dp}(\mathbf{r}_m, \mathbf{r}_n, \mathbf{r}_p, \mathbf{r}_q) = \frac{3 a_p^3 m^2 \omega \omega_z^2}{2\pi \hbar} \sqrt{\frac{\pi \hbar}{2 m \omega_z}}\\
\times e^{-\frac{3m\omega}{2\hbar} \mathbf{d}^2} \delta_{mq}\delta_{np} \delta_{<mn>}.
\end{aligned}
\end{equation}
Plugging this back into \cref{app.J}, one gets
\begin{equation}
\begin{aligned}
\tilde{J}_{ijkl} =& \frac{1}{2} \sum_{<m,n>} \phi_i^{*}(\mathbf{r}_m) \phi_j^{*}(\mathbf{r}_n)\\
\times &\frac{3 a_p^3 m^2 \omega \omega_z^2}{2\pi \hbar} \sqrt{\frac{\pi \hbar}{2 m \omega_z}}e^{-\frac{3m\omega}{2\hbar} \mathbf{d}^2}
\phi_k(\mathbf{r}_n) \phi_l(\mathbf{r}_m),
\end{aligned}
\end{equation}
and comparison with \cref{JTilde} yields
\begin{equation} \label{Vap}
V = \frac{3 a_p^3 m^2 \omega \omega_z^2}{2\pi \hbar} \sqrt{\frac{\pi \hbar}{2 m \omega_z}}e^{-\frac{3m\omega}{2\hbar} \mathbf{d}^2}.
\end{equation}

%
\end{document}